\documentclass{aastex}

\newcommand{\fluxunit}{ergs cm$^{-2}$ s$^{-1}$}

\shortauthors{Wang et al.}

\begin{document}

\title{172 ks Chandra Exposure of the LALA Bo\"{o}tes Field:\\ X-ray Source Catalog}

\author{J. X. Wang\altaffilmark{1}, S. Malhotra\altaffilmark{2}, 
J. E. Rhoads\altaffilmark{2}, M. J. I. Brown\altaffilmark{3}, 
A. Dey\altaffilmark{3}, T. M. Heckman\altaffilmark{1}, 
B. T. Jannuzi\altaffilmark{3}, C. A. Norman\altaffilmark{1,2}, 
G. P. Tiede\altaffilmark{4}, and P. Tozzi\altaffilmark{5}}
\begin{abstract}
We present an analysis of a deep, 172 ks $Chandra$ observation
of the Large Area Lyman Alpha Survey (LALA) Bo\"{o}tes field, 
obtained with the Advanced CCD Imaging Spectrometer (ACIS-I) on the 
$Chandra$ X-ray Observatory. This is one of the deepest $Chandra$
images of the extragalactic sky; only the 2 Ms CDF-N and 1 Ms CDF-S 
are substantially deeper.
In this paper we present the X-ray source catalog obtained from this
image, along with an analysis of source 
counts, and optical identifications.  
The X-ray image is
composed of two individual observations obtained in 2002, and
reaches 0.5 -- 2.0 and 2.0 -- 10.0 keV flux limits of
1.5 $\times$ 10$^{-16}$ and 1.0 $\times$ 10$^{-15}$ \fluxunit,
respectively, for point sources near the aim point.
A total of 168 X-ray sources were detected: 160 in the 0.5 -- 7.0
keV band, 132 in the 0.5 -- 2.0 keV band, and 111 in the 2.0 -- 7.0 
keV band.
The X-ray source counts were derived and compared with those from
other $Chandra$ deep surveys; the hard X-ray source density of
the LALA Bo\"{o}tes field is 33\% higher than that of CDF-S
at the flux level of 2.0 $\times$ 10$^{-15}$ 
\fluxunit, confirming the field-to-field variances of the hard band source 
counts reported by previous studies. 
The deep exposure resolves  $\gtrsim$ 72\% of the 2.0 -- 10.0 
keV X-ray background. 

Our primary optical data are $R$-band imaging from NOAO Deep Wide-Field 
Survey (NDWFS), with limiting magnitude of $R$ = 25.7 (Vega, 3$\sigma$,
4\arcsec\ diameter aperture). We have found optical counterparts for 152 of the 168
$Chandra$ sources (90\%); 144 of these are detected on the $R$-band image,
and 8 have optical counterparts in other bands (either $B_W$,$V$,$I$,$z'$). 
Among the $R$-band non-detected sources, not more than 11 of them can possibly be
at $z >$ 5, based on the hardness ratios of
their X-ray emission and nondetections in bluer bands ($B_W$,$V$).
The majority ($\sim$ 76\%) of the X-ray sources are found to have 
log($f_X$/$f_R$) within 0.0$\pm$1, which are believed to be AGNs.
Most of the X-ray faint/optically bright sources (log($f_X$/$f_R$) $<$ 
-1.0) are 
optically extended, which are low-z normal galaxies
or low luminosity AGNs. There is also a population of sources which are
X-ray $overluminous$ for their optical magnitudes (log($f_X$/$f_R$) $>$ 1.0),
which are harder in X-ray and are probably obscured AGNs.
\end{abstract}

\keywords{catalogs --- galaxies: active --- galaxies: high-redshift ---
X-rays: diffuse background --- X-rays: galaxies}

\altaffiltext{1}{Johns Hopkins University, Charles and 34th Street, Bloomberg center, Baltimore, MD 21218; jxw@pha.jhu.edu, heckman@pha.jhu.edu, norman@stsci.edu.}
\altaffiltext{2}{Space Telescope Science Institute, 3700 San Martin Drive, Baltimore, MD 21218; san@stsci.edu, rhoads@stsci.edu.} 
\altaffiltext{3}{National Optical Astronomy Observatory, 950 North Cherry Avenue, Tucson, AZ 85719; mbrown@noao.edu, dey@noao.edu, jannuzi@noao.edu.}
\altaffiltext{4}{Department of Physics and Astronomy, Bowling Green State University, Bowling Green, OH 43403; tiede@astro.ufl.edu}
\altaffiltext{5}{Istituto Nazionale di Astrofisica (INAF) - Osservatorio Astronomico, 
Via G. Tiepolo 11, 34131 Trieste, Italy; tozzi@ts.astro.it}
\section {Introduction}
A new era of X-ray astronomy has begun with
the launch of the $Chandra$ $X$-$Ray$ $Observatory$ on 1999 July 23, 
thanks to
its very high sensitivity, broad energy range and high angular resolution
(Weisskopf et al. 2002). The two deepest X-ray surveys
ever conducted, 2 Ms $Chandra$ Deep Field North (Brandt et al. 2003, Alexander
et al. 2003) and 
1 Ms $Chandra$
Deep Field South (Giacconi et al. 2002; Rosati et al. 2001), were obtained
using the Advanced CCD Imaging Spectrometer detector (ACIS;
Garmire et al. 2003) on $Chandra$ 
$X$-$Ray$ $Observatory$. These two surveys are $\sim$ 50 times more 
sensitive than the deepest pre-$Chandra$ observations in the soft X-ray band 
(0.5 -- 2.0 keV, e.g., Hasinger et al. 1998) and greater than 100 times more 
sensitive than those deepest pre-$Chandra$ observations in the hard X-ray 
band (2.0 -- 10.0 keV, e.g., Ueda et al. 1999; Fiore et al. 1999).
With these surveys, the ``diffuse'' X-ray background discovered four
decades ago (Giacconi et al. 1962) has been almost entirely resolved 
into discrete sources (i.e. $>$ 90\% in the soft band, and $\sim$ 80\%
in the hard band).

Many other deep X-ray blank-sky surveys
from $Chandra$ (Stern et al. 2002a; Yang et al. 2003; Mushotzky et al. 2000;
Manners et al.
2003) and XMM (Hasinger et al. 2001) were also performed in the past few years.
Along with multi-band observations of the detected X-ray sources, these deep
surveys have brought us many interesting results and more important science 
issues to be addressed.
These include 
the large scale structures
from 2D (Yang et al. 2003) and 3D (Gilli et al. 2003; Barger et al. 2002)
analyses, type 2 QSOs (Norman et al.
2002; Stern et al. 2002b), very high redshift X-ray selected Active Galactic 
Nuclei (AGN, e.g., Barger et al. 2003), and much more.

Here we present 
a new deep (172 ks) $Chandra$ ACIS exposure, obtained originally for 
the follow-up of
Ly$\alpha$ sources from the Large Area Lyman Alpha (LALA) survey's
Bo\"{o}tes field. 
The Large Area Lyman Alpha (LALA) Survey (Rhoads et al. 2000, 2003;
Rhoads \& Malhotra 2001, Malhotra \& Rhoads 2002) was designed to search for Lyman
$\alpha$ emitters at high redshifts through narrowband imaging.  The
survey comprises two primary fields, 36\arcmin $\times$36\arcmin\ 
in size each, located in Bo\"{o}tes (at
14:25:57 +35:32 J2000.0) and in Cetus (at 02:05:20 -04:55 J2000.0).
Both fields were chosen to be inside the
large areas of the NOAO Deep Wide-Field Survey (NDWFS), which is a deep
optical and IR
($B_WRIJHK$) imaging survey of 18 ${\rm deg}^2$ of the sky with the
primary goal of studying the evolution of large-scale structure from
$z<5$
(Jannuzi and Dey 1999; Brown et al. 2003; Jannuzi et al. 2004, in preparation).
Five broadband optical images ($B_W$, $R$, $I$ from NDWFS,
and $V$, $z'$ as part of LALA) and eight narrowband images are used to
search for Lyman-$\alpha$ emitters at z $\sim$ 4.5, 5.7, and 6.5 respectively.
The X-ray image presented here was
originally obtained to investigate the X-ray properties of the detected high
redshift Ly$\alpha$ emitters.  This study was presented by Malhotra et
al. (2003), where we placed stringent upper limits on the typical
X-ray flux of Ly$\alpha$ sources and conclude that AGN (obscured or
otherwise) cannot constitute a dominant portion of the Ly$\alpha$
source population.

This $Chandra$ exposure (172 ks) is among the deepest yet obtained by $Chandra$
of the extragalactic sky; only the CDF-N and CDF-S are substantially
deeper.  In this paper we present a full catalog of the detected X-ray sources,
along with an analysis of the X-ray source counts, and the $R$-band
magnitudes (or 3$\sigma$ upper limits) for their optical
counterparts. To study the nature of these sources, spectroscopic follow-up 
observations for these sources are under way. 

The present paper is structured as follows: we present the X-ray observations
and data reduction in Section 2, source detection and catalog in
Section 3, LogN-LogS in section 4, and optical identifications in Section 5.
Our conclusions and summary are presented in Section 6.

\section{X-ray Observations and Data Reduction}
An 178 kilo-second exposure, composed of two individual observations,
was obtained using the Advanced CCD Imaging Spectrometer (ACIS) on the
{\it Chandra X-ray Observatory\/} in the very faint (VFAINT) mode.  The 
first observation, with 120 ks exposure, was taken on 2002 April 16-17 
({\it Chandra\/} Obs ID 3130). The second observation, with 58 ks exposure,
was taken on 2002 June 9 (Obs ID 3482).
All four ACIS-I chips and ACIS-S2, ACIS-S3 chips were used, with the
telescope aimpoint centered on the ACIS-I3 chip for each exposure. The
aimpoint of Obs ID 3130 is 14:25:37.791 +35:36:00.20 (J2000.0\footnote{Coordinates throughout this paper are J2000.}), and
the aimpoint of Obs ID 3482 is 14:25:37.564 +35:35:44.32,
16$\arcsec$ away from that of Obs ID 3130.
The Galactic column density N$_H$ towards our field is 1.15 $\times$ 10$^{20}$ 
cm$^{-2}$ (Hartmann et al. 1996). 
Due to their large off-axis angle during the observations, 
the ACIS-S chips have worse spatial
resolution and lower effective area relative to the ACIS-I chips.
In this paper, data from any ACIS-S CCD were therefore ignored.

\begin{figure}
\plotone{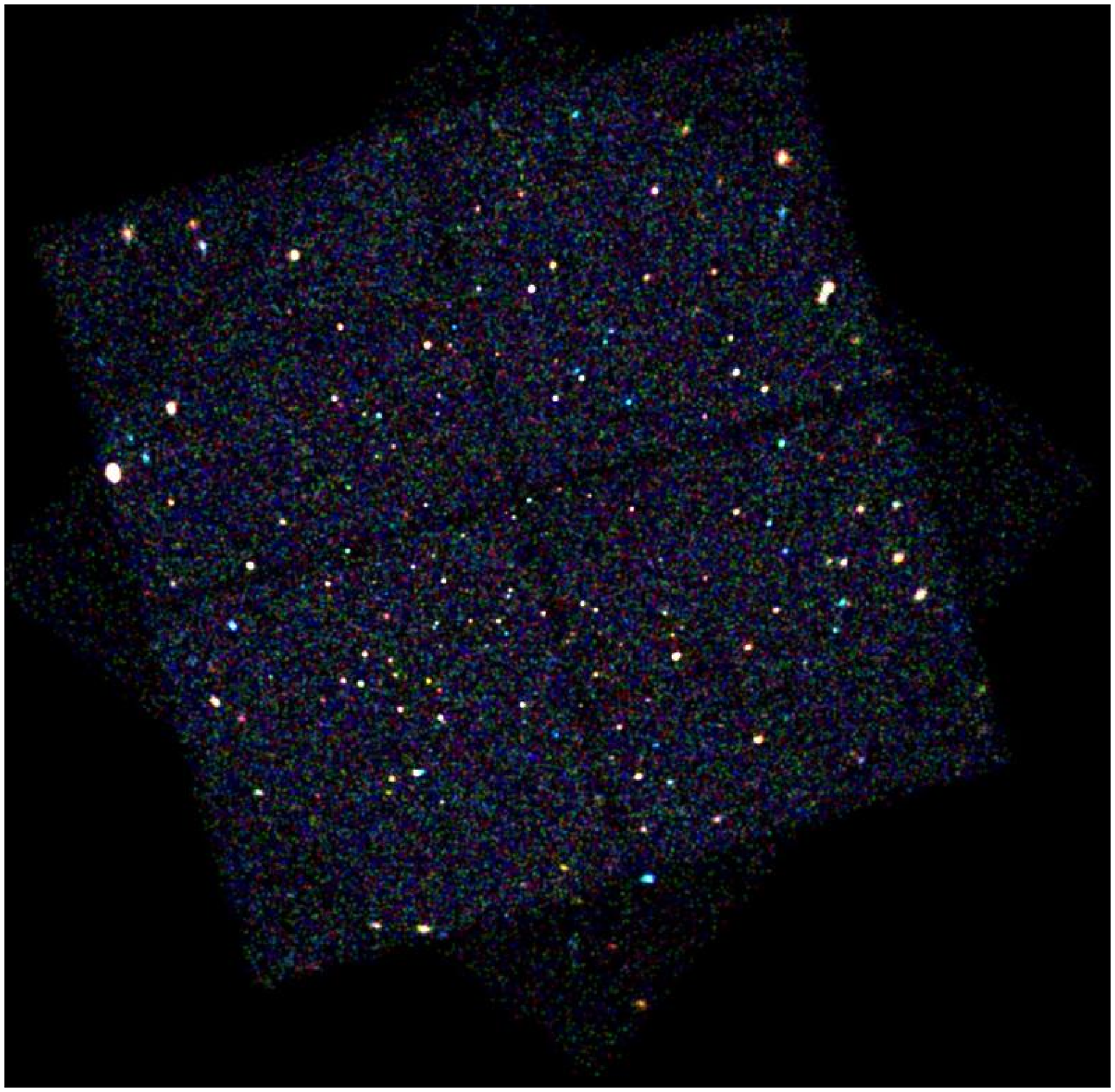}
\caption{False-color X-ray image (25$\arcmin\times$25\arcmin) of the LALA Bo\"{o}tes Field, composed
from 172 ks $Chandra$ exposure.
North is to the top, and east is to the left. The image was obtained 
combining three energy bands: 0.3 -- 1.0 keV, 1.0 -- 2.0 keV, 2.0
-- 7.0 keV (red, green and blue respectively). The color intensity
is derived directly from the counts and has not been corrected
for vignetting.
}
\label{color}
\end{figure}

The data were reduced and analyzed using the $Chandra$ Interactive
Analysis of Observations (CIAO) software (version 2.2.1, see
http://asc.harvard.edu/ciao). 
The data were reprocessed to clean the ACIS particle background 
for very faint mode observations, and then filtered to include only the
standard event grades 0,2,3,4,6\footnote{see http://asc.harvard.edu/proposer/POG/html/ACIS.html$\#$sec:GRADES}. 
All bad pixels and columns were also
removed, not only from the photon events files, but also when calculating
the effective exposure maps. The high background time intervals were
manually removed by checking the total event rates.
The total net exposure time is 172 ks (120 ks from Obs ID 3130, and 52 ks from 
Obs ID 3482). The offset between the astrometry of the two observations 
was obtained by registering the X-ray sources showing up in both exposures.
The two event files were merged after correcting the small offset (0.3\arcsec),
and the combined data have the same coordinate system as Obs ID 3130.
We present the color composite X-ray image of the combined exposure 
in Fig. \ref{color}.

\section{Source Detection and Catalog}

Three images were extracted from the combined event file for source 
detection: a soft image (0.5 -- 2.0 keV), a hard image (2.0 -- 7.0 keV) 
and a total image (0.5 -- 7.0 keV). The hard and total bands were cut 
at 7 keV since the effective area of $Chandra$ decreases above this energy, 
and the instrumental background rises, giving a very inefficient detection 
of sky and source photons. 
The WAVDETECT program (Freeman et al. 2002), which is included with the 
CIAO software package, was run on the extracted
images. A probability threshold of 1 $\times$ 10$^{-7}$ (corresponding
to 0.5 false sources expected per image), and wavelet scales of
1,2,4,8,16 pixels (1 pixel = 0.492\arcsec) were used. 
A total of 168 X-ray sources were detected: 160 in the total band
(0.5 -- 7.0 keV), 132 in the soft band (0.5 -- 2.0 keV), and 111
in the hard band (2.0 -- 7.0 keV). We present the catalog of the
detected sources in Table 1. Note that the number of total detected
X-ray sources will be increased to 196 if we use a probability
threshold of WAVDETECT (1 $\times$ 10$^{-6}$), however, considering
the increased number of possible false detections ($\sim$ 5 expected
per image), we prefer to publish a conservative catalog with fewer
false detections.
We also tried to search for X-ray sources with scales much larger than the PSF
by running WAVDETECT with wavelet scales of 32,64,128 pixels on the images.
However, we didn't detect any new source beyond those
in the catalog. Below we give explanations to the
columns in Table 1.

Column(1)-(4): the source ID, IAU name, right ascension and declination
respectively. The IAU name for the sources is CXOLALA1 JHHMMSS.s+DDMMSS.
The positions
were determined by WAVDETECT. Whenever possible, we use positions
derived in the soft band, which has the best spatial resolution
among the three bands. For sources which are not detected in the
soft band, we use total band positions instead if available, and
hard band positions were quoted for those sources detected only in the
hard band.

Column(5): 3 $\sigma$ uncertainties of the centroid positions directly given 
by WAVDETECT. 

Column(6)-(8): the net counts in the soft, hard and total bands.
The counts were calculated using circular aperture photometry.
For each source, we defined a source region which is a circle
centered at the position given in column (3) and (4), with 
radius R$_s$ set to the 95\% encircled-energy radius of $Chandra$
ACIS PSF at the source position. 
R$_s$ varies in the range of 2$\arcsec$ to 15$\arcsec$ from the
center to the edge of the field.
Source photons were then extracted from
the regions, and the local background was extracted from an 
annulus with outer radius of 2.4 $\times$ R$_s$ and inner radius
of 1.2 $\times$ R$_s$, after masking out nearby sources.
The aperture correction ($\times$ 1/0.95) was applied to the
source counts. The derived
net counts and 1 $\sigma$ Poisson uncertainties (e.g., Gehrels 1986)
are given for each source in each band.

Column(9): indication of source detection. We mark the source 
with ``T'', ``S'' and ``H'' for source detected in total, soft
and hard band respectively. Multiple letters are used for
sources detected in more than one band. For example, ``TS''
means detections in both the total band and the soft band, but
non-detected in the hard band.

\begin{figure}
\plotone{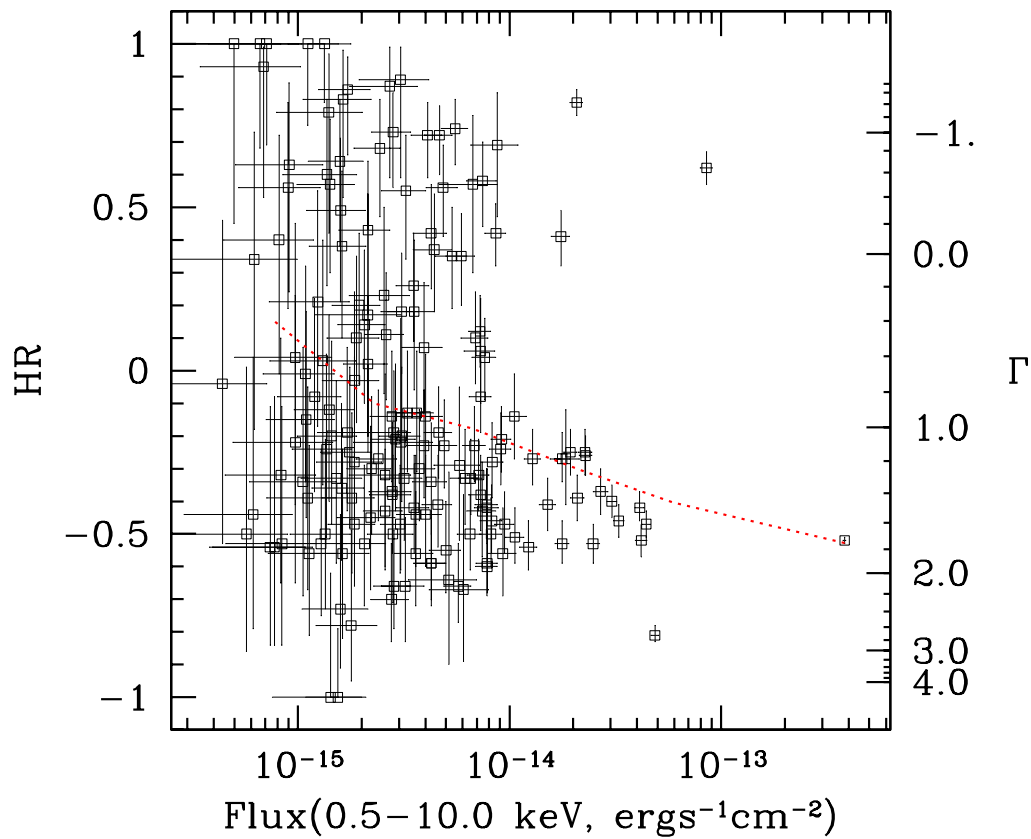}
\caption{Hardness ratios HR = $(H-S)/(H+S)$ (left ordinate axis) of X-ray sources vs their
full band (0.5 -- 10.0 keV) X-ray fluxes. The photon indices $\Gamma$ of
the power-law spectra which could reproduce the observed hardness
ratios are given along the right ordinate (see text for details). The dotted line connects
the averaged hardness ratios for detected sources within different flux
bins.}
\label{hr}
\end{figure}

Column(10): hardness ratio defined as $HR$ = (H-S)/(H+S), where H and S 
are the vignetting-corrected net counts in the hard and soft band respectively.
The hardness ratios vs 0.5 -- 10.0 keV band X-ray fluxes for the detected
X-ray sources are plotted in Fig. \ref{hr}.
Assuming a power-law spectrum with
the Galactic HI column density (1.15 $\times$ 10$^{20}$ cm$^{-2}$),
the observed hardness ratio can be converted to the photon index $\Gamma$ of the
spectrum, which is also presented in the figure.
As presented in earlier surveys, harder sources are seen
at fainter fluxes, most of which are believed to be obscured AGNs.

\begin{figure}
\plotone{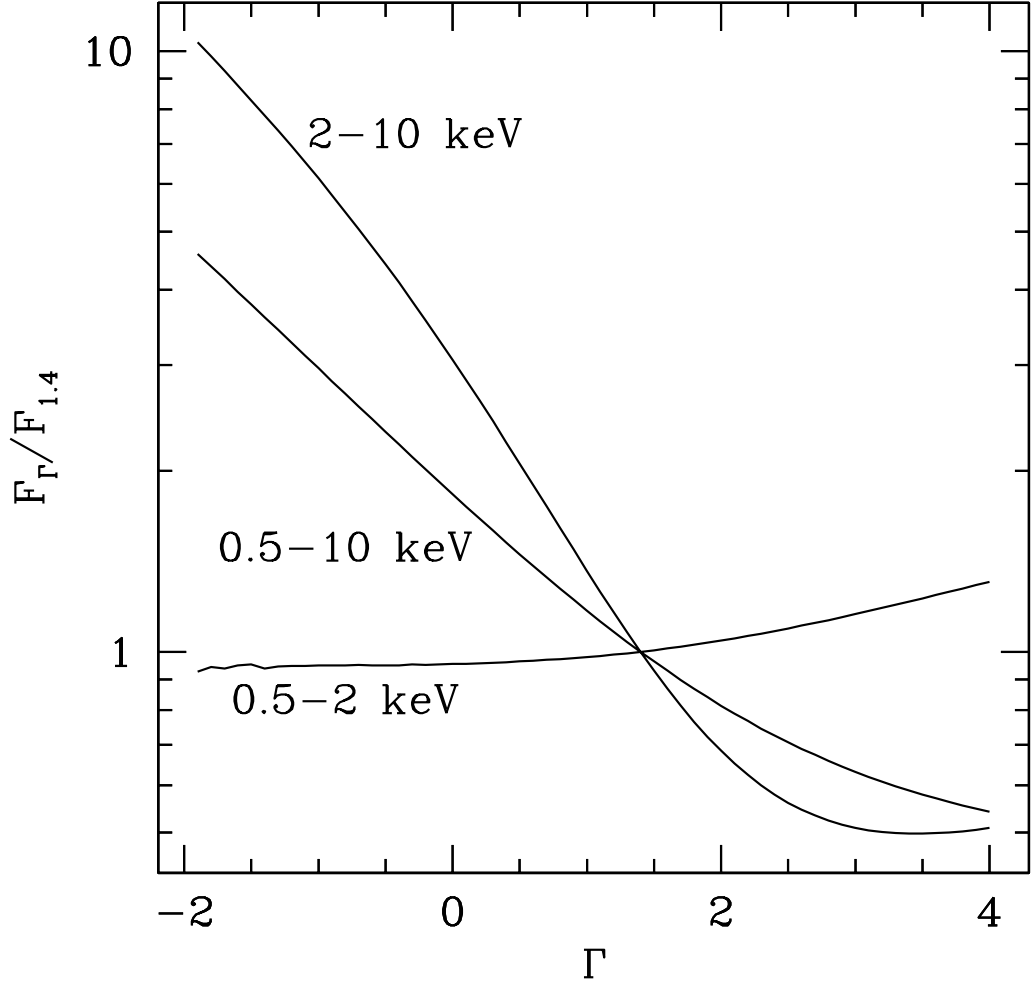}
\caption{Conversion factors to calculate three band fluxes assuming a
power-law spectrum with photon index different from $\Gamma$ = 1.4. 
}
\label{cf}
\end{figure}

Column(11)-(13): X-ray fluxes (Galactic absorption corrected) of 
three bands in the unit of 10$^{-15}$ \fluxunit. 
A power-law spectrum with the Galactic column density was assumed
to calculate the conversion factors from net counts to X-ray fluxes.
The photon index of the power-law was chosen at $\Gamma$ = 1.4,
which was also used in Giacconi et al. (2002) and Stern 
et al. (2002a). Three band net count rates were calculated by
dividing the net counts in column 6-8 by the effective exposure
time\footnote 
{The effective exposure time was calculated through multiplying 172 ks
by the ratio of the exposure map at the aim point to the value of the
exposure map averaged within the extraction region for each source.
Such a correction was done in each band separately.}
at each source position in each band, and then
converted into X-ray fluxes of 0.5 -- 10.0 keV, 0.5 -- 2.0
keV, and 2.0 -- 10.0 keV respectively. 
These settings make
our results directly comparable with those from other surveys.
The conversion factors used were 1.25 $\times$ 10$^{-11}$ ergs cm$^{-2}$
count$^{-1}$ from the 0.5 -- 7.0 keV band observed count rates to the
Galactic absorption corrected X-ray fluxes in the band 0.5 -- 10.0 keV,
4.67 $\times$ 10$^{-12}$ ergs cm$^{-2}$ count$^{-1}$ from the 0.5 -- 2.0
keV band count rates to the 0.5 -- 2.0 keV band
fluxes, and 2.96 $\times$ 10$^{-11}$ ergs cm$^{-2}$
count$^{-1}$ from the 2.0 -- 7.0 keV band count rates to the
2.0 -- 10.0 keV band fluxes.
Note the total band (0.5 -- 10.0 keV) flux is not equal to the 
sum of the soft and hard band fluxes if the actual photon index
differs from 1.4. 
To calculate the fluxes assuming a power-law spectrum with
different photon index $\Gamma$, the conversion factors 
F$_{\Gamma}$/F$_{1.4}$ for different bands are needed
(Fig. \ref{cf}). We can see from the figure that the soft band 
(0.5 -- 2.0 keV) flux is not sensitive to the photon index $\Gamma$, 
while the hard (2.0 -- 10.0 keV) and total (0.5 -- 10.0 keV) band fluxes 
correlate with it strongly.  
One thus has to be careful while using these fluxes.

Column(14)-(16): The offsets of the detected optical counterparts 
from the X-ray source positions ($\Delta\alpha$ =
RA$_R$ - RA$_X$, $\Delta\delta$ = Dec$_R$ - Dec$_X$),
and the $R$-band AUTO magnitudes
(Kron-like elliptical aperture magnitudes, Bertin \& Arnouts 1996) with 1 $\sigma$ 
uncertainties\footnote{The uncertainties of the
magnitudes are direct output from SExtractor, without including
the uncertainty of the $R$-band zeropoint.}. For sources which are not
detected in $R$-band, 3$\sigma$ upper limits of the magnitudes 
are given when available. See section 5 for details.

Column(17): The FWHM (full-width half-maximum) of the optical counterparts
in NDWFS $R$-band image (see section 5).

\section {X-ray source counts}
\begin{figure}
\plotone{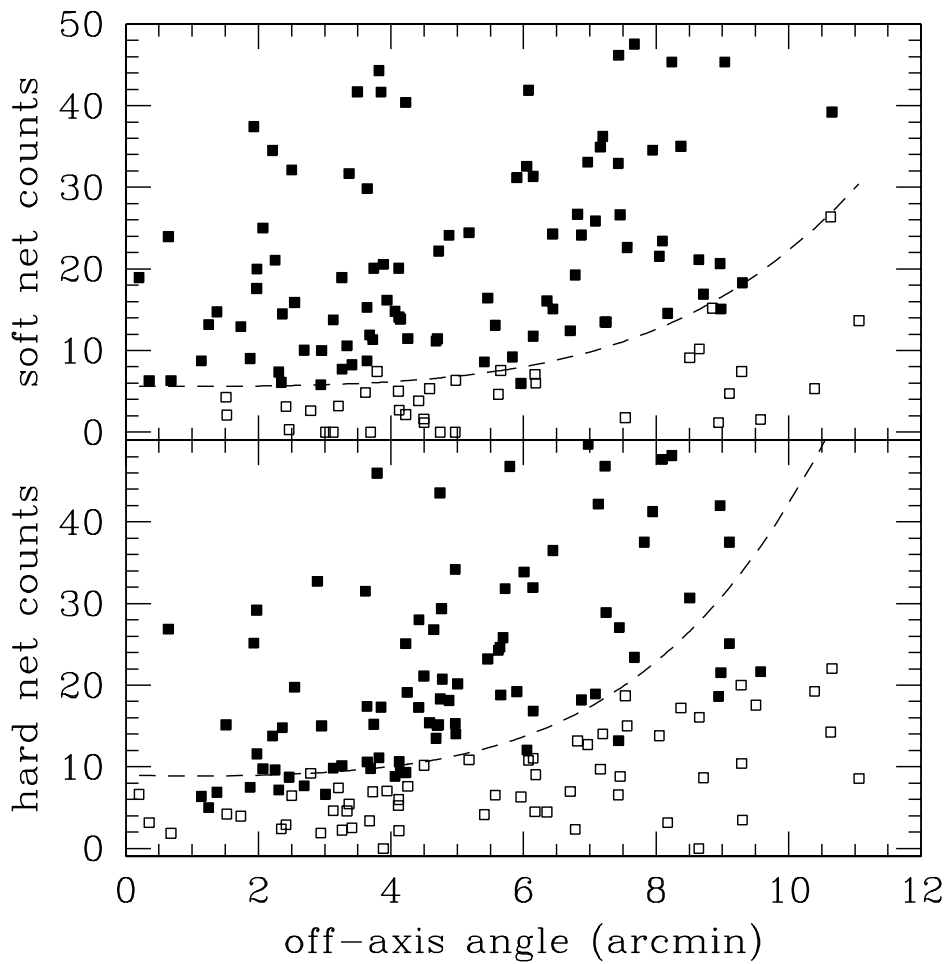}
\caption{Soft (0.5 -- 2.0 keV) and hard (2.0 -- 7.0 keV) band net counts
vs off-axis angle. The two dashed lines are the threshold we chose
to build complete samples for LogN-LogS calculation.
The net counts are derived from X-ray photometry. Thus for each
source we can give net count for each band, whether it is
detected in the band or not.
In the upper
panel, the filled squares are sources detected in the soft band,
and open ones are those not detected in the soft band. 
In the lower panel,  the filled squares are sources detected in the hard band,
and open ones are those not detected in the hard band.
}
\label{cutoff}
\end{figure}

\begin{figure}
\plotone{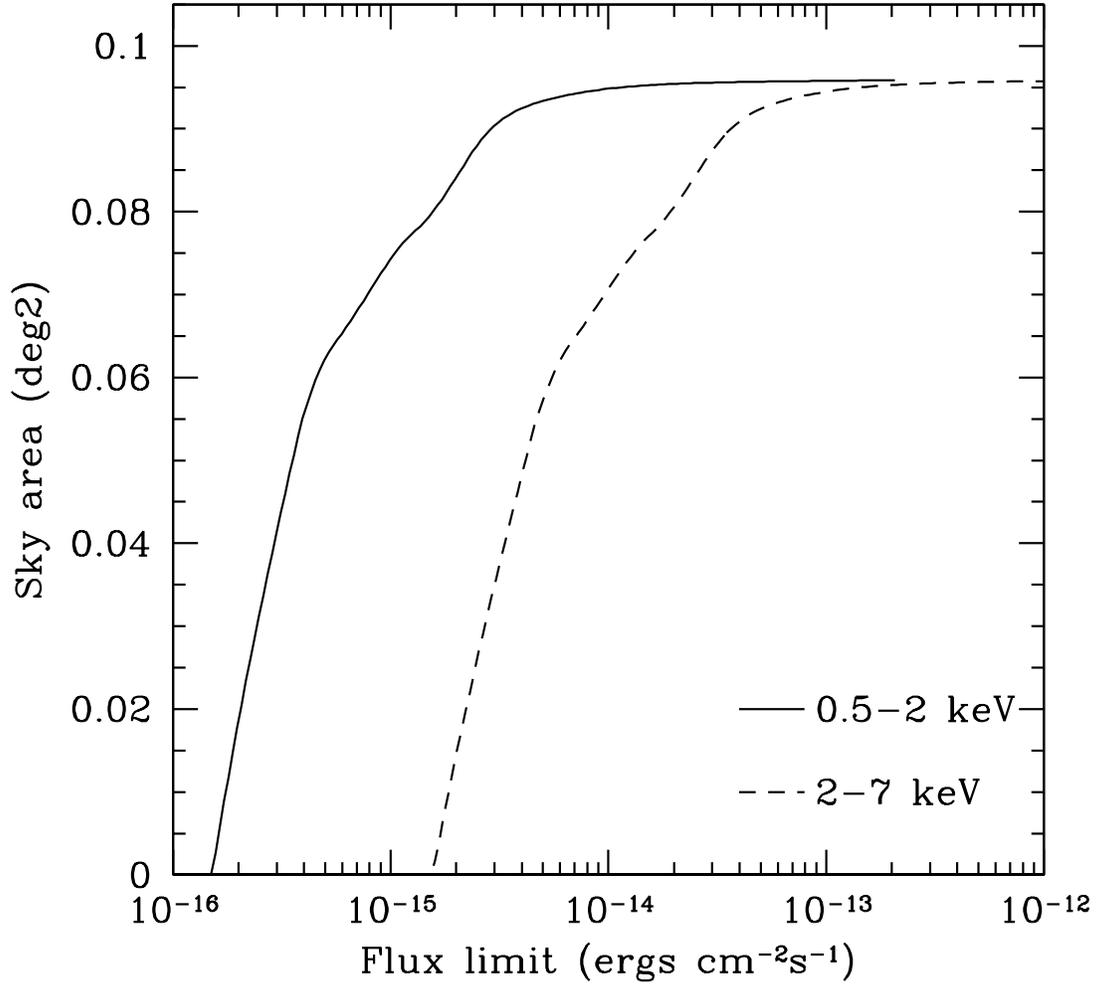}
\caption{Sky coverage in soft (0.5 -- 2.0 keV) and hard (2.0 -- 7.0 keV) 
bands, as a function of flux limit.}
\label{sc}
\end{figure}

\begin{figure}
\plotone{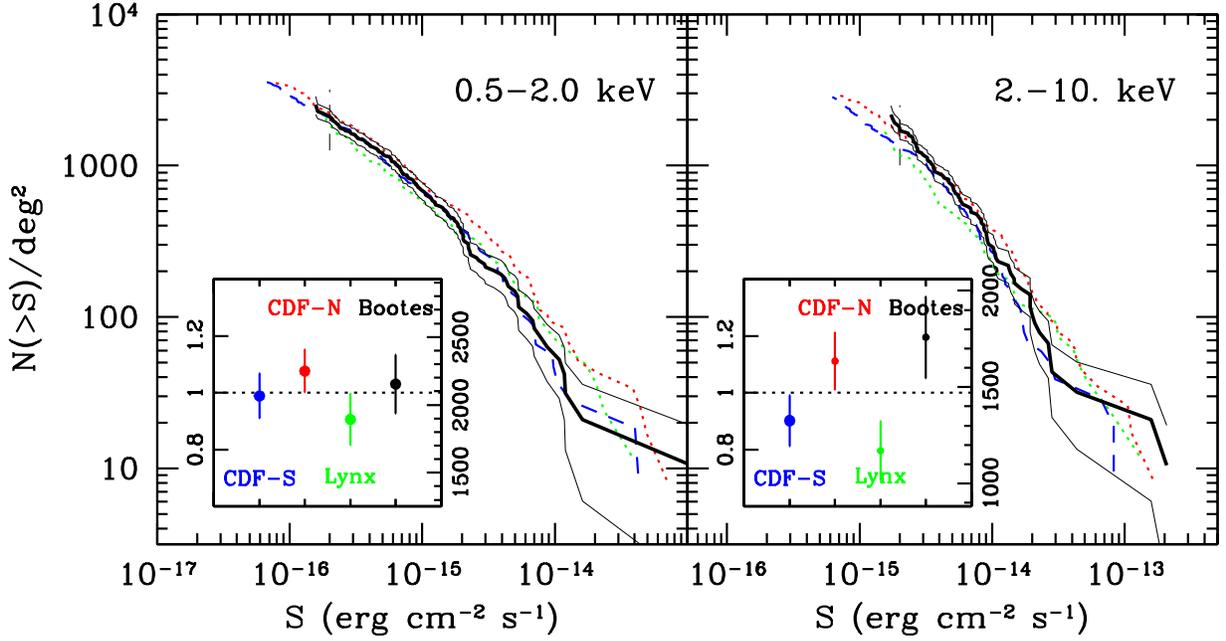}
\caption{LogN-LogS in the soft (0.5 -- 2.0 keV) and hard  (2.0 --
 10.0 keV) bands from 172ks $Chandra$ observations of LALA Bo\"{o}tes
 field. 
 Data are plotted as thick solid lines with two additional thin solid 
lines enclosing 1$\sigma$ 
 Poisson uncertainties. 
LogN-LogS from CDF-N, CDF-S and Lynx field are also plotted (see text
for details).
The inserts show the X-ray source densities and 1$\sigma$
uncertainties from the four
fields at the faint end of our 172 ks $Chandra$ exposure
(2.0 $\times$ 10$^{-16}$ \fluxunit\ in 0.5 -- 2.0 keV band, and 2.0 $\times$
10$^{-15}$ \fluxunit\ in the 2.0 -- 10.0 keV band).
The dashed lines in the inserts are the average source densities from
the four fields, and the scales of the inserts are 0.6 to 1.4 times
the average values.
}
\label{lognlogs}
\end{figure}

In order to calculate the cumulative source counts, N($>$S),
one need count the number of sources with fluxes $>$S, and also 
compute the summed sky in the field where these sources can be detected 
(the sky coverage).
In Fig. \ref{cutoff}, we present the soft and hard band net counts vs off-axis
angle for faint sources (with net counts $<$ 50). Detected 
and non-detected sources (detected in other bands but having net counts 
from X-ray photometry) are displayed with different symbols.
It's clear from the figure that the X-ray sources with larger off-axis angles
need more net counts to be detected because of the larger PSF size.
This indicates that the detection limits of net counts vary with the PSF sizes.
Two dashed lines (net counts = A + B $\times$ PSF$^2$ respectively) are added 
to 
Fig. \ref{cutoff}, with the parameters A and B chosen for both bands by visual 
inspection to exclude all non-detected sources and include maximum number of detected 
sources. We can see that although some detected sources are located
below the threshold dashed lines, they are mixed up with these non-detected
sources, i.e., the detection is incomplete below the lines, and the
sky coverage for these sources is thus not available.
In this paper, we use only these sources with net counts 
C $>$ A + B $\times$ PSF$^2$ to calculate LogN-LogS\footnote{A = 5.3, 8.2 and
B = 0.026, 0.053 for the soft and hard band respectively.}.
Using the above cutoffs, the sky coverages 
are derived and presented in Fig. \ref{sc}.
The derived LogN-LogS for both soft and hard band are presented in
Fig. \ref{lognlogs}.
A maximum-likelihood power-law was used to fit the slope
of the LogN-LogS in each band. For the 0.5 -- 2.0 keV band we find
\begin{equation}
N (> S) = 340 ({ S \over {2 \times 10^{-15}\ {\rm ergs}\ {\rm cm^{-2}}\ {\rm s^{-1}} } })^{-0.78}
\end{equation}
And for the 2.0 -- 10.0 keV band, we find:
\begin{equation}
N (> S) = 1790 ({ S \over {2 \times 10^{-15}\ {\rm ergs}\ {\rm cm^{-2}}\ {\rm s^{-1}} } })^{-1.16}
\end{equation}

The above procedures of calculating LogN-LogS were also run on the
1 Ms CDF-N data (Brandt at al. 2001), 1 Ms CDF-S data (Rosati et al. 2002),
and 184.7 ks Lynx data (Stern et al. 2002a). The independently derived 
LogN-LogS from the above three fields
match the published ones to within 1$\sigma$, and are plotted in 
Fig. \ref{lognlogs}
for comparison. The source densities (N$>$S) and 1$\sigma$ uncertainties
at the faint end
(2.0 $\times$ 10$^{-16}$ \fluxunit\ in 0.5 -- 2.0 keV band, and 2.0 $\times$
10$^{-15}$ \fluxunit\ in the 2.0 -- 10.0 keV band)
are plotted in the inserts.
In the soft band, there is no significant difference among the source counts
from the four deep surveys.
In the hard band, obvious fluctuations of the source counts are seen at the faint
end: LALA Bo\"otes field has the highest source density at 2.0 $\times$ 
10$^{-15}$ \fluxunit\, which is 33\%
higher than that of CDF-S,
while CDF-N is 23\% higher than CDF-S and Lynx field is 14\% lower.
Note that similar field-to-field variances of the hard band source counts
have been reported previously (see Tozzi 2001a, Cowie et al. 2002). 
This is believed to be due to the clustering of the
X-ray sources.
The non-detection of field-to-field 
variance in the soft band is also consistent with the results from previous
studies (Yang et al. 2003; Tozzi et al. 2001a), indicating that the
soft X-ray sources are less correlated than the hard X-ray sources.

What fraction of the hard X-ray background is resolved by our
deep 172 ks $Chandra$ imaging?
In the range of 1.7-100 $\times$ 10$^{-15}$ \fluxunit, 
the integrated hard X-ray flux density in the 2.0 -- 10.0 keV band is 1.2 
$\times$ 10$^{-11}$ \fluxunit\ deg$^{-2}$. Note for sources bright than 10$^{-13}$ \fluxunit,
the integrated hard X-ray flux density from our survey is 0.4 $\times$ 10$^{-11}$ 
\fluxunit\ deg$^{-2}$, consistent with the value derived by della Ceca et al.
(1999) from a wider area ASCA survey.
The total integrated hard X-ray flux density we obtained is 
1.6 $\times$ 10$^{-11}$ \fluxunit\ deg$^{-2}$, 
down to 1.7 $\times$ 10$^{-15}$ \fluxunit\ in
2.0 -- 10.0 keV band.
This is equal to the High Energy Astronomy Observatory 1
(HEAO 1) value (Marshall et al. 1980), but 10\%-30\% lower than the more
recent determinations from ASCA and $BeppoSAX$ (e.g.,
Ueda et al. 1999; Vecchi et al. 1999). We conclude that our 172 $Chandra$
deep imaging resolves $\gtrsim$ 70\% of the 2.0 -- 10.0 keV X-ray background,
and the main uncertainty comes from the value of total X-ray background itself.

\section{Optical Identifications}
\subsection{Optical Images}

Our deep Chandra pointing is within the
NDWFS Bo\"otes subfield NDWFS J1426+3531 (which is 35\arcmin$\times$35\arcmin\ in size) 
centered at RA 14:26:00.8, DEC +35:31:32.0 (J2000).
Five broadband optical images ($B_W$, $V$, $R$, $I$, and $z'$)
are available to search for the optical counterparts of the X-ray sources.
Three of these ($B_W$, $R$, and $I$)\footnote{The $B_W$ and $I$ band images
are available from the NOAO Science Archive 
(http://www.archive.noao.edu/ndwfs/). 
The $R$-band image, which is deeper than the currently released version,
will be available from the NOAO Science Archive
within 12 months.}
are from the NOAO Deep Wide Field Survey (NDWFS; Jannuzi \& Dey 1999), 
while the remaining two ($V$ and $z'$) were obtained as part of LALA. 
The limiting Kron-Cousins system magnitudes (3$\sigma$,
4\arcsec\ diameter aperture)
of the NDWFS images are $B_W < 26.5$,
$R<25.7$, and $I<25.0$. The corresponding limits in the LALA broadband
data are $V<25.9$ and $z'<24.6$ (where the $z'$ limit is on the AB
system). 
Our primary optical image for identifications in this paper is the $R$-band
image from NDWFS, which is substantially deeper than the others.
All the optical images were obtained using the Kitt
Peak National Observatory Mayall 4m telescope and the Mosaic-1 camera
(Muller et al. 1998; Jannuzi et al. 2004, in preparation).  
These images were produced following the procedures described in
version 7.01 of "The NOAO Deep Wide-Field Survey MOSAIC Data
Reductions"\footnote{http://www.noao.edu/noao/noaodeep/ReductionOpt/frames.html.}.
A general description of the software used is provided by Valdes (2002)
and the complete details of the NDWFS data reduction will be
provided by Jannuzi et al. (2004).  

\subsection{Optical Counterparts}
SExtractor package (Bertin \& Arnouts 1996) V2.1.0 was run on the
NDWFS deep $R$-band image to generate the $R$-band source catalog. 
The registration of the X-ray to optical coordinates was done by
cross-correlating the X-ray and the above $R$-band catalogs (e.g.,
see Giacconi et al. 2002, Stern et al. 2002a).
Average shifts in (RA$_R$ - RA$_X$, Dec$_R$ - Dec$_X$) of (0.1\arcsec, 
-0.2\arcsec) were found from the NDWFS $R$-band to X-ray imaging, but
no obvious rotation or plate-scale effects were discovered.
After correcting the average shifts, we matched the optical and X-ray source 
catalogs using a 1.5\arcsec\ radius aperture for host identifications, which
was generally used by other $Chandra$ surveys (Giacconi et al. 2002,
Stern et al. 2002a).
The 3$\sigma$ X-ray positional uncertainties in Table 1
are used if larger than 1.5\arcsec.

Optical counterparts for 144 of the 168 X-ray sources were detected
in the $R$-band image down to 3$\sigma$. One of them (No. 122)
has multiple possible optical counterparts. 
Calculations show that the possibility of having one optical source
down to 3$\sigma$ in a 1.5\arcsec\ circle by chance is 10\%.
This means that at least 90\% of the optical counterparts we found should correspond
to the X-ray sources, and at worst we might have $\sim$ 14 false matches.

\begin{figure}
\plotone{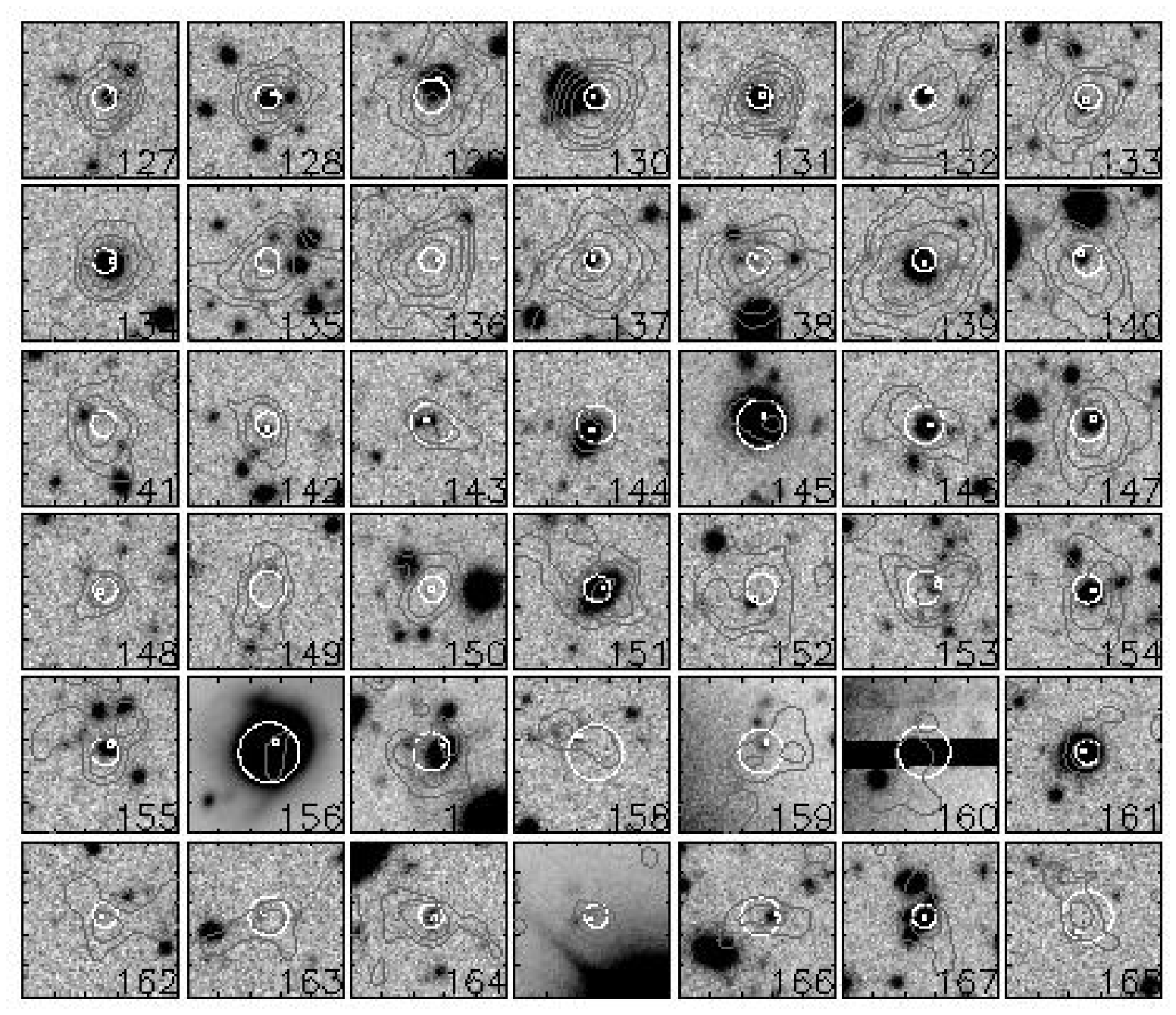}
\caption{
Optical cutouts (20$\arcsec\times$20$\arcsec$) for the 168 X-ray sources, 
overlapped by black X-ray isointensity contours (at 3,5,10,20 and 100 $\sigma$ 
above the local background).
The small white
box indicates an optical counterpart candidate, and the circle in the center
indicates the radius used to search counterparts.
North is to the top, and east is to the left.
}
\label{cutout}
\end{figure}

We present the offsets from the X-ray source positions ($\Delta\alpha$ = 
RA$_R$ - RA$_X$, $\Delta\delta$ = Dec$_R$ - Dec$_X$),
the derived $R$-band AUTO magnitudes (Kron-like elliptical aperture magnitudes, Bertin \& Arnouts 1996) for all optical counterparts 
in Table 1, and the optical cutouts in Fig. \ref{cutout}.
IRAF package "radprof" was used to measure the radial profiles of these
counterparts in NDWFS $R$ image, and we present the measured FWHM
(full-width half-maximum) in Table 1. These sources which have FWHM larger
than the average seeing of the NDWFS $R$ image (FWHM = 1.16\arcsec) were
examined visually. We conclude that it is safe to consider all the
bright sources ($R$ $<$ 23) with FWHM $\ge$ 1.3\arcsec\ as optically resolved.
The only exception is source No. 144, which is too close to another bright
source, thus we can not tell if it is resolved or not.
The total number of optically resolved sources is 15.

The search for optical counterparts was extended to $B_W$, $I$, $V$ and 
$z'$ band
for the 24 X-ray sources without $R$-band optical counterparts,
which brought us optical counterparts for 8 more X-ray sources.
The positions of these counterparts are also given in Table 1 (by listing
RA - RA$_X$, Dec - Dec$_X$). 
The limiting magnitude of the $R$-band
image was given as upper limits for 22 of the X-ray sources.
Note source No. 37 and 160 are overlapped by bleeding trails in the
NDWFS $R$ image, and the upper limits of their $R$-band magnitudes
are not available.

The FIRST survey (Faint Images of the Radio Sky at Twenty-cm, Becker, White,
\& Helfand 
1995) detection limit in our field is 0.96 mJy/beam.
Radio counterparts for 4 of our X-ray sources were found, which are
source No. 1, 66, 91 and 141, with integrated flux density (mJy) of 6.02, 1.24,
2.17 and 8.30 respectively.
The first 3 of the 4 radio sources are resolved 
R$<$21 red galaxies, which are some of the brightest optical
counterparts to X-ray sources in our catalog. The remaining source is an
R-band non-detection more than 10" from any R$<$21 galaxies and is
presumably at z$>$1.

\subsection{X-ray-to-optical flux ratio}
\begin{figure}
\plottwo{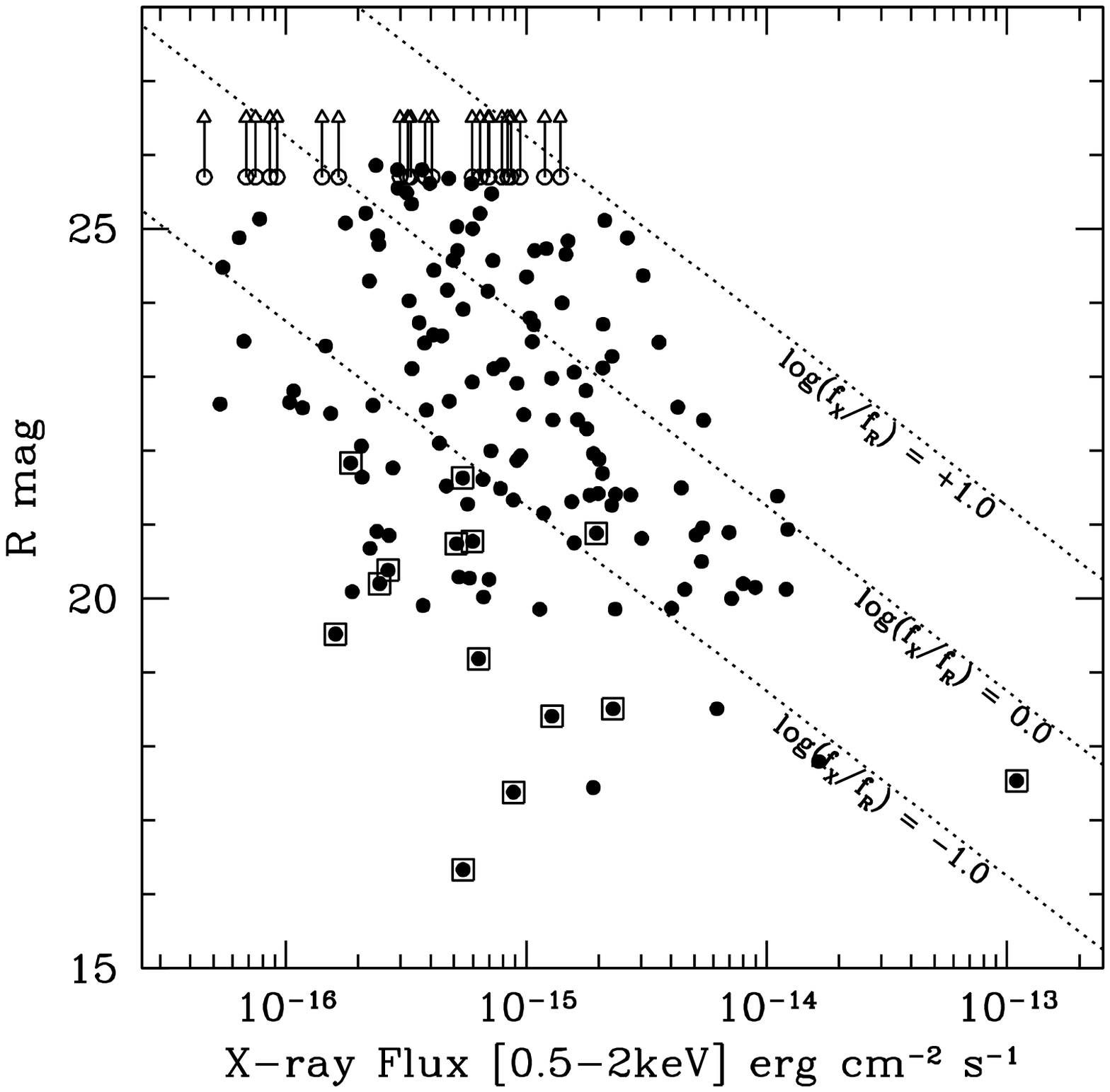}{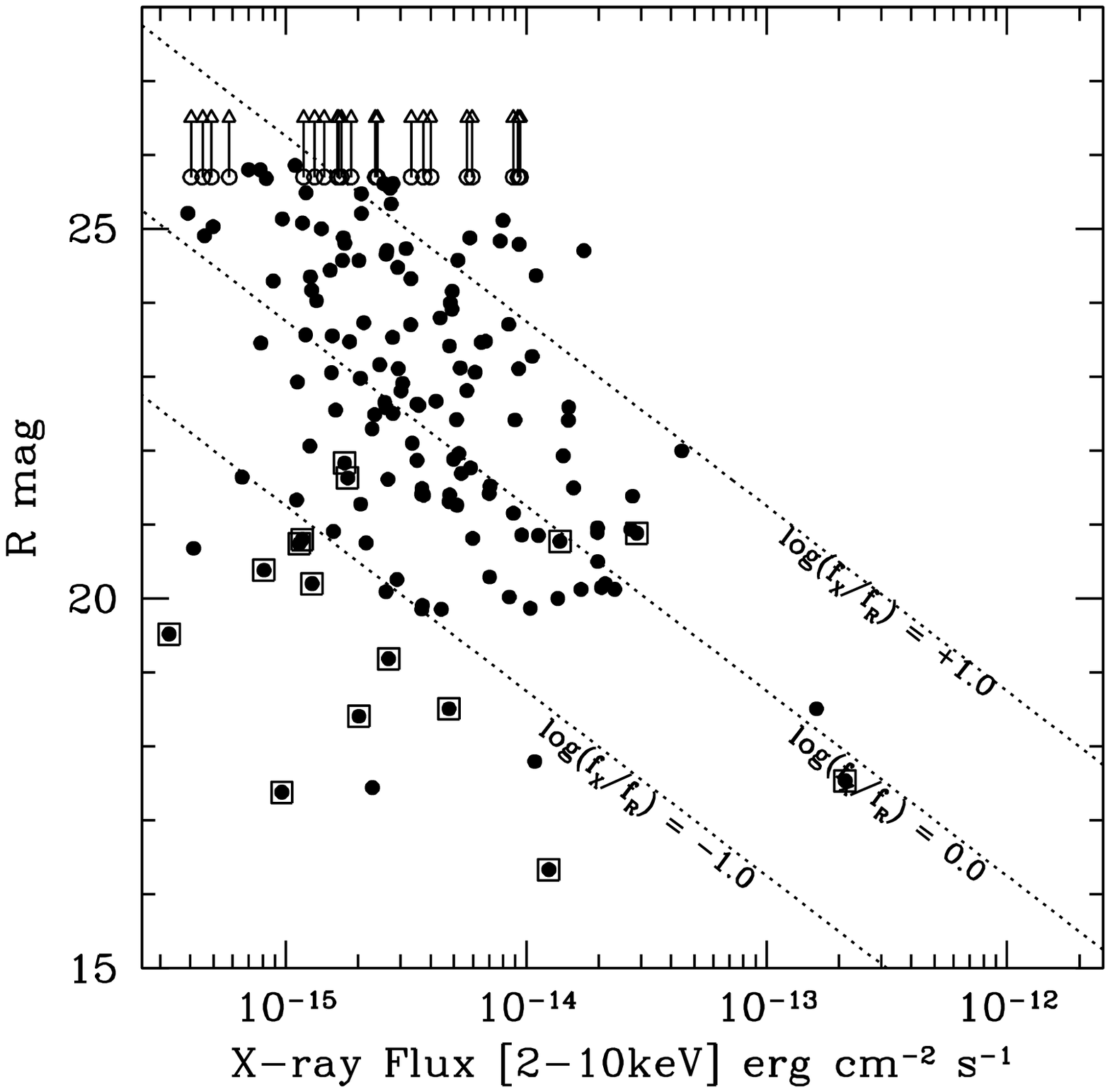}
\caption{
Optical $R$-band magnitudes of the X-ray detected sources vs their soft 
(0.5 -- 2.0 keV) and hard (2 -- 10 keV) X-ray fluxes. 3$\sigma$ upper 
limits are plotted for sources without $R$-band counterparts.
Dotted lines show location of constant X-ray-to-optical flux ratio 
log($f_X$/$f_R$) of +1, 0, and -1.
Dots enclosed by squares are optically extended sources.
}
\label{xr}
\end{figure}

\begin{figure}
\plotone{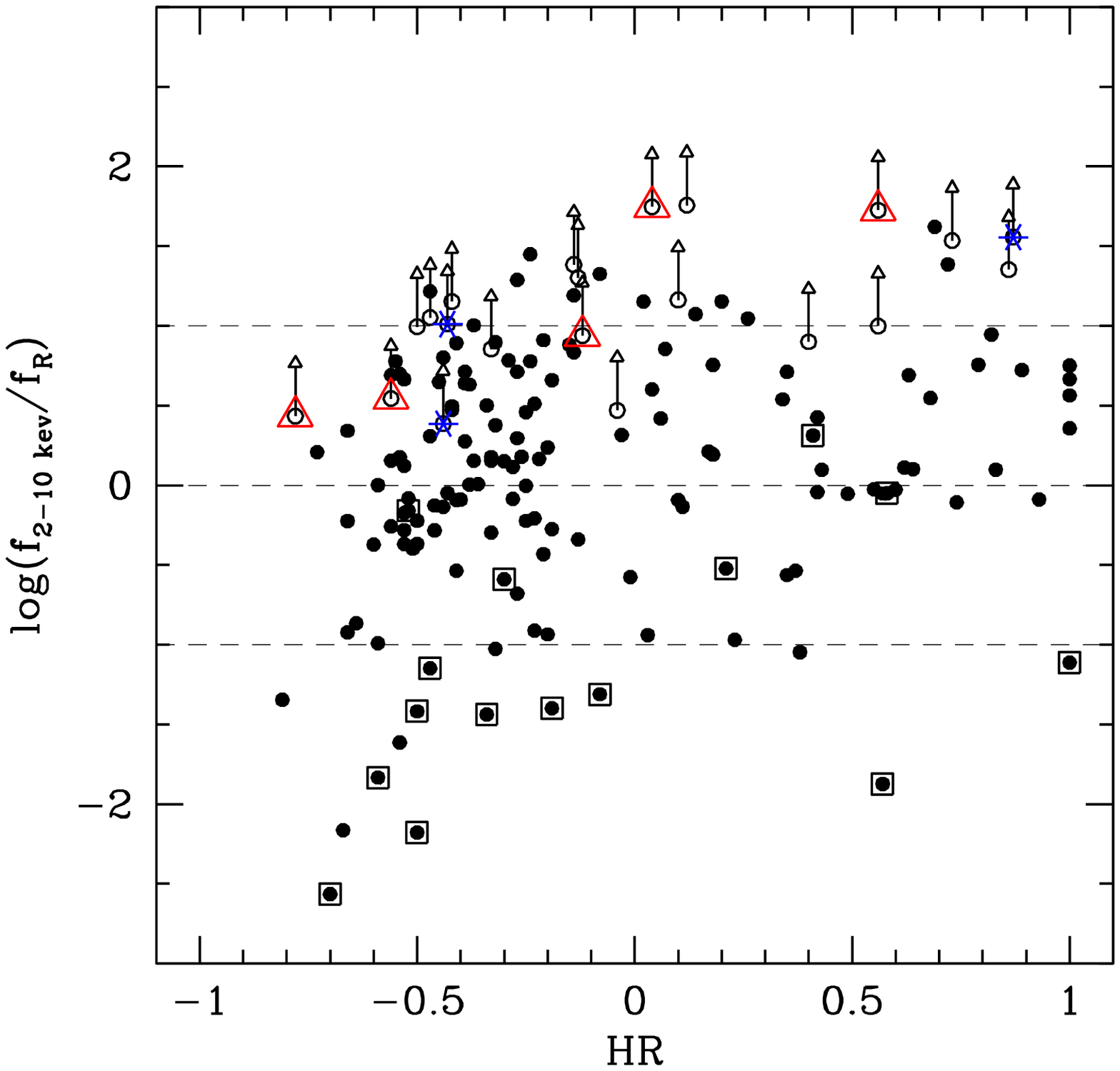}
\caption{X-ray-to-optical flux ratio log($f_{2-10keV}$/$f_R$) vs X-ray 
hardness ratio HR. Lower limits of log($f_{2-10keV}$/$f_R$) are plotted
for sources without $R$-band counterparts.
Dashed lines show location of constant X-ray-to-optical flux ratio
of +1, 0, and -1.
Dots enclosed by squares are optically extended sources, open
circles enclosed by triangles are $R$-band nondetected sources which
show up in redder bands ($I$,$z'$), and stars are bluer bands ($B_W$,
$V$) detected only.
}
\label{xrhr}
\end{figure}

In Fig. \ref{xr} we plot the X-ray-to-optical flux ratio of our X-ray
sources (in soft and hard band respectively), and in Fig. \ref{xrhr}
we plot the log($f_X$/$f_R$) in the hard band vs the hardness ratio of
X-ray emission.
For the 22 X-ray sources without $R$-band counterparts detected, the
lower limits of log($f_X$/$f_R$) were plotted, which shows a consistent
distribution with that of the optically faint ($R$ $>$ 24) sources.
From the figures we can see that the majority ($\sim$ 76\%) of the X-ray sources fall
within log($f_{2-10keV}$/$f_R$) = 0.0$\pm$1, which is typical of AGNs. 

There is a significant population ($\sim$ 10\%) of X-ray faint/optically bright
sources (log($f_{2-10keV}$/$f_R$) $<$ -1.0), most of which are extended in the $R$
band optical image and X-ray soft (see Fig. \ref{xrhr}). 
These sources should be nearby,
bright normal galaxies (see Tozzi et al. 2001b; Barger et al. 2001; 
Hornschemeier at al. 2001). Other optically extended sources falling
within log($f_{2-10keV}$/$f_R$) = 0.0$\pm$1 or X-ray hard are most likely to be low 
redshift Seyfert galaxies, or low luminosity AGNs.

\begin{figure}
\plotone{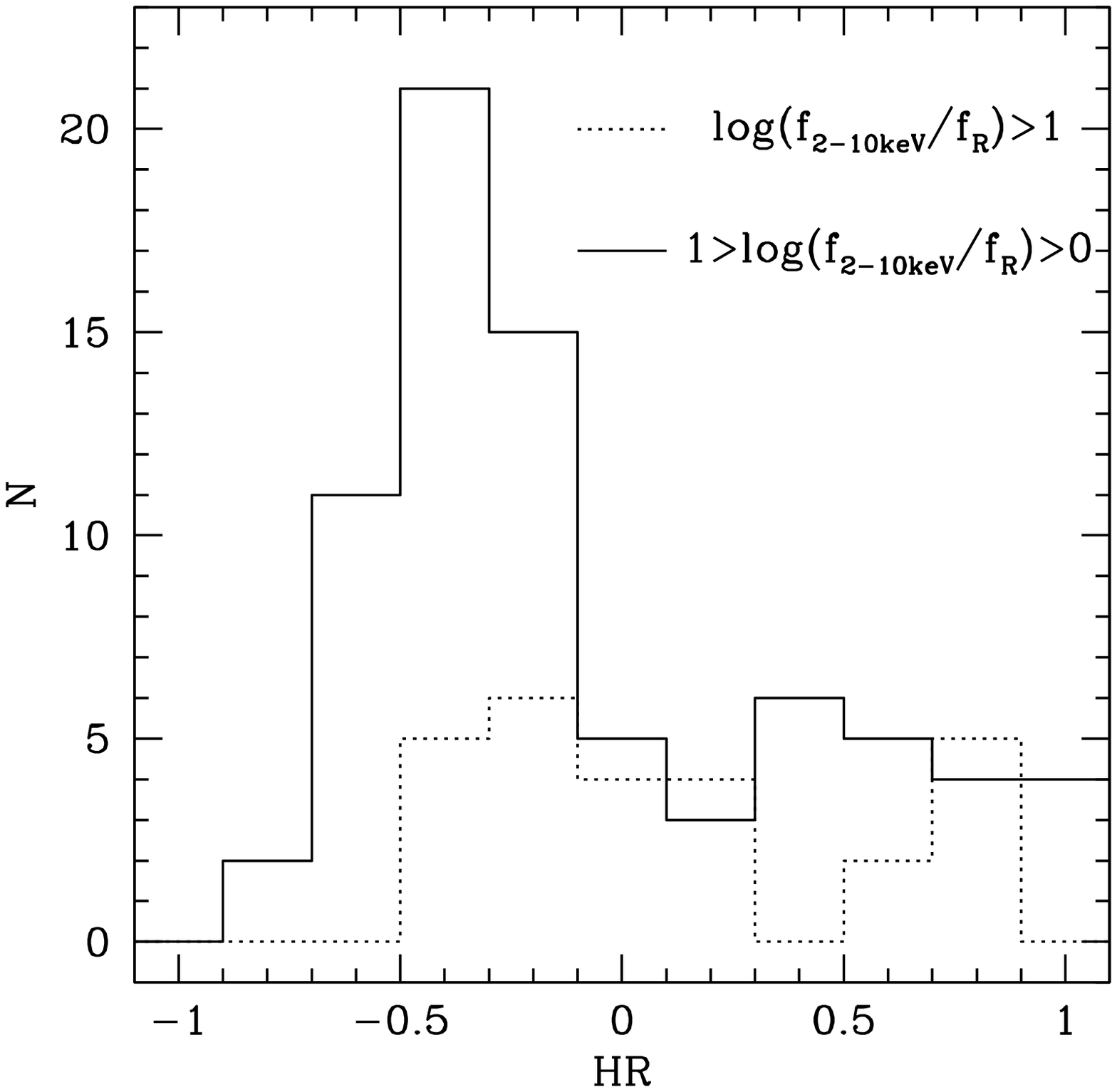}
\caption{
Histogram distribution of X-ray hardness ratio for the optically faint
X-ray bright sources (log($f_{2-10keV}$/$f_R$) $>$ 1), comparing with
these sources with 1.0 $>$ log($f_{2-10keV}$/$f_R$) $>$ 0.0.
The two distributions are distinguishable according to the K-S test
at the 98.7\% significance level.
}
\label{hist}
\end{figure}

In addition, we also have a population of sources which are
X-ray $overluminous$ for their optical magnitudes (log($f_{2-10keV}$/$f_R$) $>$ 1.0),
especially in the hard band, including most
of the 22 $R$-band nondetected X-ray sources. 
We plot in Fig. \ref{hist} the histogram distribution of the hardness ratios of
the optically faint X-ray bright sources (log($f_{2-10 keV}$/$f_R$) $>$ 1.0), 
comparing with these sources with lower X-ray-to-optical flux ratio
(1.0 $>$ log($f_{2-10 keV}$/$f_R$) $>$ 0.0).
The optically faint X-ray bright sources are obviously 
harder in X-ray, suggesting that most of them are obscured AGNs,
for which high column density absorption shields both the optical
and the soft X-ray emission.

\subsection{$R$-band nondetections}

There are 22 X-ray sources which are not detected in the $R$-band
down to $R$ = 25.7; 5 of them are detected in redder bands ($I$,$z'$);
and 3 of them in bluer bands ($B_W$, $V$); 14 X-ray sources have no
optical counterparts found in any band, which corresponds
to a sky surface density of 145 deg$^{-2}$.
The integrated hard X-ray (2.0 -- 10.0 keV) flux 
density from these 22 sources is 0.14 $\times$ 10$^{-11}$ 
\fluxunit\ deg$^{-2}$, down to a flux limit of 1.7 $\times$ 
10$^{-15}$ \fluxunit. This contributes 6$\sim$9\% of the total
2.0 -- 10.0 keV band X-ray background.

The 19 $R$-band nondetected sources, which are not detected in
bluer bands ($B_W$, $V$), are possible candidates for $z$ $>$ 5 quasars, 
because of the absorption from
the Lyman transitions of hydrogen along our line of sight.
However, our calculations indicate that even for a Compton thick AGN (with
intrinsic photon power-law index $\Gamma$ = 2.0 and absorption column density
N$_H$ = 10$^{24}$ cm$^{-2}$), the hardness ratio HR is expected to be $<$ 0
at $z$ $>$ 5. Thus 8 of them with harder
X-ray spectra (HR $>$ 0) can not be at $z$ $>$ 5.
In Fig. \ref{xrhr}, we also marked the 5 X-ray sources without
$R$-band counterparts, but showing up in redder bands ($I$,$z'$).
Preliminary results from photometric redshift analyses
show that the two softest ones of the 5 are at photo-z $\sim$ 4.3, and
the other 3 which have harder X-ray spectra at photo-z $\sim$ 1-2.
This agrees well with the above argument that high redshift objects
should be soft in X-ray.
Excluding these sources with HR $>$ 0, 
there are 11 $R$-band nondetections left, which can be possibly at $z$ $>$ 5.

Assuming a constant AGN density up to $z$ = 10, Gilli, Salvati \& Hasinger (2001) predicted 
a $z$ $>$ 5 AGN density of 500 deg$^{-2}$ for a limiting flux of 
2.3 $\times$ 10$^{-15}$ \fluxunit\ in the 0.5 -- 2.0 keV band.
This model predicts $\sim$ 50 $z >$ 5 sources in our field,
which is clearly too high.
Assuming the AGN density above $z$ = 2.7 decreases by a factor of 2.7
per unit redshift as found for optical quasars (Schmidt, Schneider, \& Gunn 1995), 
Gilli et al. predicted a much lower source
density of $z$ $>$ 5 AGNs (25 deg$^{-2}$) for the same limiting flux.
It corresponds to $\sim$2 $z$ $>$ 5 objects in our field,
which is very plausible with the current data, considering that
those 11 sources are unlikely to be all at $z$ $>$ 5.
Actually, several optically faint $Chandra$ sources from other surveys
have already been confirmed to be at $z$ $<$ 5
by spectroscopic observations (e.g., see Alexander et al. 2001).

\section {Conclusions}
We present a deep, 172 ks $Chandra$ ACIS exposure of the LALA Bo\"{o}tes field.
This paper describes the details of the observations, data reduction,
source detection, LogN-LogS analysis, and presents the X-ray source catalog
along with $R$-band magnitudes of their optical counterparts.
A total of 168 X-ray sources were detected, 160 in the total band
(0.5 -- 7.0 keV), 132 in the soft band (0.5 -- 2.0 keV), and 111
in the hard band (2.0 -- 7.0 keV). Near the aim point, the detection
is down to a flux limit of 1.5 
$\times$ 10$^{-16}$ \fluxunit\ in soft (0.5 -- 2.0 keV) band, and 1.0 
$\times$ 10$^{-15}$ \fluxunit\ in hard (2.0 -- 10.0 keV) band.
LogN-LogS was compared with those from other deep surveys, and we find
obvious field-to-field fluctuations of the hard band source counts.
These fluctuations are believed to be due to the large scale clustering of the detected
X-ray sources. Calculations show that our deep imaging resolves $\gtrsim$ 70\% of 
the X-ray background in 2.0 -- 10.0 keV band. Optical counterparts for
90\% of the X-ray sources were found from deep optical images.
Among the $R$-band non-detected sources, 11 of them can be possibly
at $z >$ 5, based on the hardness ratios of
their X-ray emission and the fact that they are not detected in bluer
bands ($B_W$,$V$).
Spectroscopic follow-up observations of these counterparts are being undertaken
and will be present in future paper.

\acknowledgements 
This work was supported by the CXC grant GO2-3152X and the National Optical
Astronomy Observatory which is operated by the Association of Universities 
for Research in Astronomy (AURA), Inc.  under a cooperative agreement with 
the National Science Foundation.
We would like to thank the NDWFS team, and in particular Melissa Miller 
(NOAO), who assisted in the reduction of the $R$-band image used 
extensively in this paper. We would like to thank N. Brandt for 
providing the CDF-N LogN - LogS. We also thank the referee for a
prompt and helpful report.

\clearpage

\newpage

\begin{figure}
\figurenum{7}
\plotone{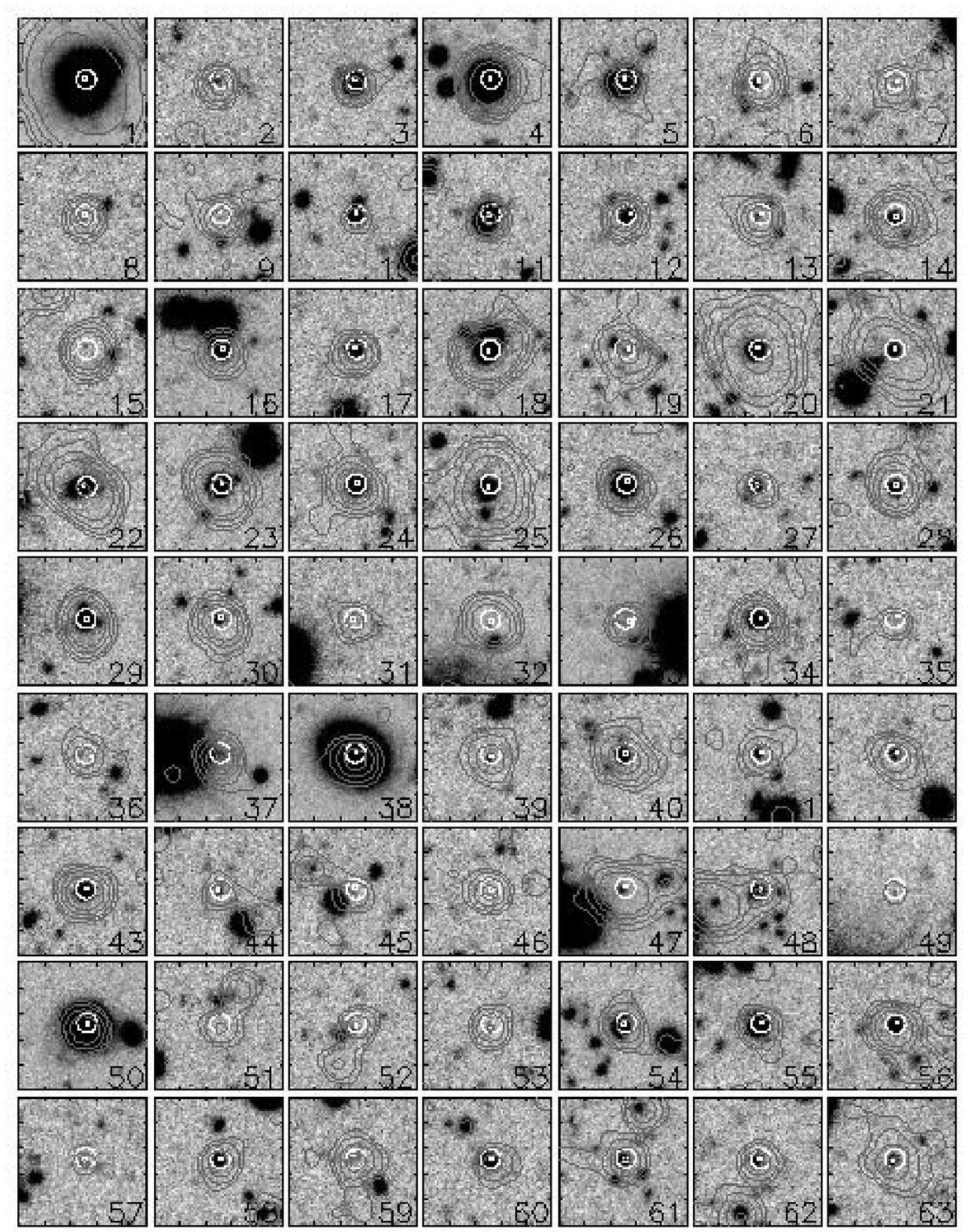}
\caption{}
\end{figure}

\begin{figure}
\figurenum{7}
\plotone{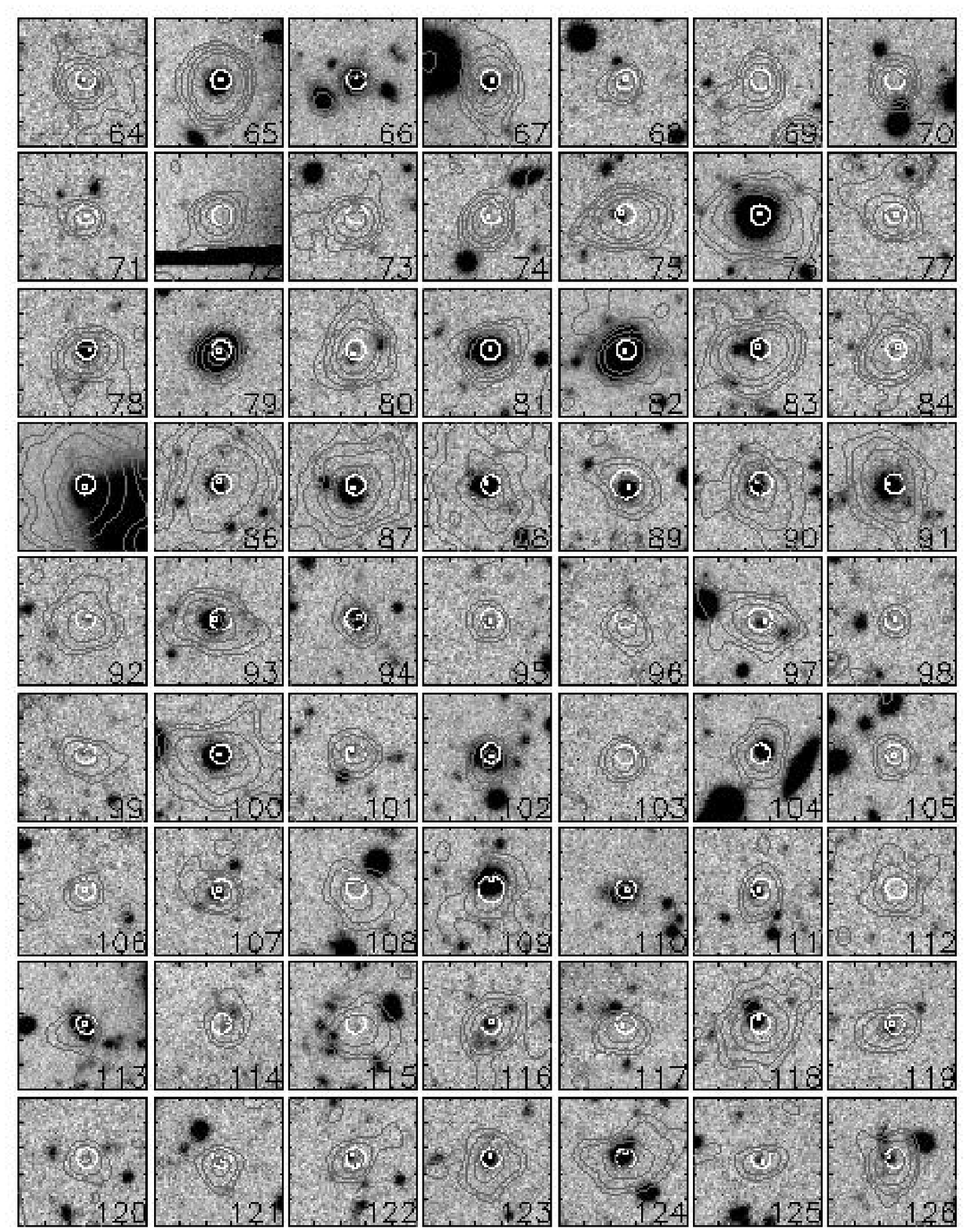}
\caption{}
\end{figure}

\begin{deluxetable}{r@{~~~}l@{}r@{~~~}r@{~~~}r@{~~~}r@{~~~}r@{~~~}c@{~~~}r@{~~~}r@{~~~}r@{~~~}c@{~~~}c@{~~~}l@{~~~}c@{~~~}}
\rotate
\tabletypesize{\tiny}
\tablewidth{0pc}
\tablecaption{$Chandra$ Sources in LALA Bo\"{o}tes Field\label{common}}
\tablehead{
\colhead{ID} & \colhead{\space\space CXOLALA1\space\space\space\space\space\space\space\space\space RA(J2000)\space\space  Dec(J2000)} & \colhead{Err} & \colhead{Tot. Cts.} & \colhead{Soft Cts.} & \colhead{Hard Cts.} & \colhead{det} & \colhead{HR} & \colhead{S$_{0.5-10}$} & \colhead{S$_{0.5-2}$} & \colhead{S$_{2-10}$} &\colhead{$\Delta\alpha(\arcsec)$} &\colhead{$\Delta\delta(\arcsec)$} &\colhead{$R$} &\colhead{FWHM(\arcsec)}}

\startdata
1	& J142620.3+353708\space\space\space14:26:20.36\space\space\space35:37:08.7	& 0.2\arcsec	& $4722.9^{+71.9}_{-70.8}$ & $3607.8^{+62.8}_{-61.8}$ & $1115.2^{+35.7}_{-34.6}$ &	TSH	& -0.52$^{+0.01}_{-0.01}$	& $382.34$ & $109.67$ & $212.18$ 	& 0.3 & -0.1 & $17.54\pm{0.00}$ 	& 1.3 \\
2	& J142555.5+353528\space\space\space14:25:55.50\space\space\space35:35:28.5	& 0.5\arcsec	& $32.7^{+7.1}_{-5.9}$ & $15.3^{+5.2}_{-4.0}$ & $17.4^{+5.5}_{-4.3}$ &	TSH	& 0.07$^{+0.20}_{-0.20}$	& $3.94$ & $0.69$ & $4.92$ 	& 0.6 & -0.0 & $24.16\pm{0.12}$ 	& 1.1 \\
3	& J142547.2+353627\space\space\space14:25:47.25\space\space\space35:36:27.6	& 0.4\arcsec	& $31.6^{+6.9}_{-5.7}$ & $20.0^{+5.7}_{-4.5}$ & $11.6^{+4.6}_{-3.4}$ &	TSH	& -0.27$^{+0.21}_{-0.19}$	& $2.40$ & $0.57$ & $2.05$ 	& 0.4 & -0.1 & $21.28\pm{0.01}$ 	& 1.2 \\
4	& J142547.1+353954\space\space\space14:25:47.10\space\space\space35:39:54.9	& 0.2\arcsec	& $656.4^{+27.4}_{-26.4}$ & $593.6^{+26.1}_{-25.0}$ & $62.8^{+9.4}_{-8.3}$ &	TSH	& -0.81$^{+0.03}_{-0.02}$	& $48.63$ & $16.50$ & $10.84$ 	& 0.2 & -0.1 & $17.80\pm{0.00}$ 	& 1.1 \\
5	& J142546.4+353522\space\space\space14:25:46.48\space\space\space35:35:22.3	& 0.4\arcsec	& $16.5^{+5.5}_{-4.3}$ & $ 9.0^{+4.3}_{-3.1}$ & $ 7.5^{+4.1}_{-2.9}$ &	TSH	& -0.08$^{+0.30}_{-0.29}$	& $1.20$ & $0.25$ & $1.28$ 	& 0.3 & -0.1 & $20.20\pm{0.01}$ 	& 1.4 \\
6	& J142545.3+353449\space\space\space14:25:45.33\space\space\space35:34:49.7	& 0.3\arcsec	& $62.6^{+9.3}_{-8.2}$ & $37.4^{+7.4}_{-6.3}$ & $25.1^{+6.4}_{-5.2}$ &	TSH	& -0.19$^{+0.14}_{-0.14}$	& $4.62$ & $1.04$ & $4.38$ 	& 0.3 & -0.2 & $23.80\pm{0.08}$ 	& 1.0 \\
7	& J142543.3+353548\space\space\space14:25:43.31\space\space\space35:35:48.1	& 0.4\arcsec	& $15.1^{+5.3}_{-4.2}$ & $ 8.7^{+4.3}_{-3.1}$ & $ 6.4^{+4.0}_{-2.7}$ &	TSH	& -0.15$^{+0.33}_{-0.30}$	& $1.09$ & $0.24$ & $1.09$ 	& -0.1 & -0.4 & $25.86\pm{0.35}$ 	& 0.8 \\
8	& J142542.5+353358\space\space\space14:25:42.53\space\space\space35:33:58.2	& 0.4\arcsec	& $30.7^{+6.9}_{-5.7}$ & $21.1^{+5.8}_{-4.7}$ & $ 9.6^{+4.5}_{-3.3}$ &	TSH	& -0.37$^{+0.21}_{-0.19}$	& $2.79$ & $0.72$ & $2.06$ 	& 0.4 & -0.3 & $25.48\pm{0.28}$ 	& 0.9 \\
9	& J142539.5+353357\space\space\space14:25:39.55\space\space\space35:33:57.9	& 0.4\arcsec	& $34.8^{+7.2}_{-6.1}$ & $25.0^{+6.3}_{-5.1}$ & $ 9.8^{+4.5}_{-3.3}$ &	TSH	& -0.43$^{+0.19}_{-0.17}$	& $2.58$ & $0.69$ & $1.70$ 	& 0.2 & 0.2 & $>$25.7	& ...\\
10	& J142538.7+353618\space\space\space14:25:38.72\space\space\space35:36:18.0	& 0.4\arcsec	& $ 9.5^{+4.3}_{-3.1}$ & $ 6.3^{+3.8}_{-2.5}$ & $ 3.2^{+3.1}_{-1.7}$ &	TS	& -0.32$^{+0.42}_{-0.33}$	& $0.83$ & $0.21$ & $0.66$ 	& -0.0 & -0.0 & $21.64\pm{0.02}$ 	& 1.1 \\
11	& J142537.9+353612\space\space\space14:25:37.94\space\space\space35:36:12.0	& 0.3\arcsec	& $25.6^{+6.4}_{-5.2}$ & $18.9^{+5.6}_{-4.4}$ & $ 6.6^{+4.0}_{-2.7}$ &	TS	& -0.47$^{+0.23}_{-0.19}$	& $1.85$ & $0.51$ & $1.14$ 	& 0.4 & -0.3 & $20.74\pm{0.01}$ 	& 1.3 \\
12	& J142536.3+353634\space\space\space14:25:36.32\space\space\space35:36:34.2	& 0.3\arcsec	& $50.8^{+8.5}_{-7.3}$ & $24.0^{+6.2}_{-5.0}$ & $26.9^{+6.5}_{-5.3}$ &	TSH	& 0.06$^{+0.16}_{-0.16}$	& $7.30$ & $1.29$ & $8.96$ 	& -0.6 & -0.0 & $22.41\pm{0.03}$ 	& 1.2 \\
13	& J142534.7+353407\space\space\space14:25:34.76\space\space\space35:34:07.8	& 0.3\arcsec	& $46.8^{+8.2}_{-7.1}$ & $17.6^{+5.6}_{-4.4}$ & $29.2^{+6.7}_{-5.5}$ &	TSH	& 0.26$^{+0.16}_{-0.17}$	& $3.53$ & $0.50$ & $5.19$ 	& 0.0 & -0.1 & $24.58\pm{0.15}$ 	& 1.0 \\
14	& J142533.5+353845\space\space\space14:25:33.51\space\space\space35:38:45.6	& 0.3\arcsec	& $95.9^{+11.3}_{-10.2}$ & $63.2^{+9.3}_{-8.2}$ & $32.7^{+7.2}_{-6.0}$ &	TSH	& -0.32$^{+0.11}_{-0.11}$	& $7.15$ & $1.77$ & $5.66$ 	& -0.1 & -0.5 & $22.81\pm{0.04}$ 	& 1.1 \\
15	& J142530.7+353911\space\space\space14:25:30.71\space\space\space35:39:11.3	& 0.3\arcsec	& $95.2^{+11.2}_{-10.1}$ & $41.7^{+7.8}_{-6.7}$ & $53.5^{+8.7}_{-7.6}$ &	TSH	& 0.12$^{+0.11}_{-0.11}$	& $7.25$ & $1.19$ & $9.46$ 	& ... & ... & $>$25.7	& ...\\
16	& J142529.2+353412\space\space\space14:25:29.24\space\space\space35:34:12.2	& 0.4\arcsec	& $38.6^{+7.7}_{-6.5}$ & $32.1^{+7.0}_{-5.8}$ & $ 6.5^{+4.1}_{-2.9}$ &	TS	& -0.66$^{+0.17}_{-0.13}$	& $2.84$ & $0.88$ & $1.11$ 	& -0.1 & -0.0 & $21.33\pm{0.01}$ 	& 1.1 \\
17	& J142525.4+353622\space\space\space14:25:25.40\space\space\space35:36:22.8	& 0.5\arcsec	& $35.6^{+7.3}_{-6.2}$ & $15.9^{+5.3}_{-4.2}$ & $19.8^{+5.7}_{-4.5}$ &	TSH	& 0.11$^{+0.19}_{-0.20}$	& $2.61$ & $0.44$ & $3.35$ 	& 0.2 & -0.1 & $22.10\pm{0.02}$ 	& 1.1 \\
18	& J142520.7+353311\space\space\space14:25:20.75\space\space\space35:33:11.7	& 0.3\arcsec	& $410.9^{+22.1}_{-21.0}$ & $300.5^{+19.0}_{-17.9}$ & $110.4^{+12.1}_{-11.0}$ &	TSH	& -0.46$^{+0.05}_{-0.05}$	& $32.83$ & $8.97$ & $20.55$ 	& 0.4 & -0.3 & $20.15\pm{0.00}$ 	& 1.0 \\
19	& J142614.2+353631\space\space\space14:26:14.29\space\space\space35:36:31.5	& 1.0\arcsec	& $59.4^{+9.9}_{-8.8}$ & $46.2^{+8.4}_{-7.3}$ & $13.2^{+6.1}_{-4.9}$ &	TSH	& -0.55$^{+0.15}_{-0.13}$	& $5.01$ & $1.46$ & $2.61$ 	& -0.5 & -0.1 & $24.65\pm{0.08}$ 	& 1.1 \\
20	& J142614.2+353833\space\space\space14:26:14.21\space\space\space35:38:33.1	& 0.6\arcsec	& $558.0^{+25.7}_{-24.7}$ & $411.4^{+22.0}_{-21.0}$ & $146.6^{+14.1}_{-13.0}$ &	TSH	& -0.47$^{+0.04}_{-0.04}$	& $44.21$ & $12.24$ & $27.22$ 	& 0.5 & -0.0 & $20.93\pm{0.01}$ 	& 1.0 \\
21	& J142609.5+353213\space\space\space14:26:09.56\space\space\space35:32:13.8	& 0.7\arcsec	& $246.2^{+17.8}_{-16.7}$ & $171.2^{+14.7}_{-13.6}$ & $75.0^{+10.8}_{-9.7}$ &	TSH	& -0.39$^{+0.07}_{-0.07}$	& $20.91$ & $5.46$ & $14.97$ 	& 0.0 & -0.1 & $22.41\pm{0.03}$ 	& 1.0 \\
22	& J142607.6+353351\space\space\space14:26:07.66\space\space\space35:33:51.4	& 0.6\arcsec	& $265.1^{+18.1}_{-17.0}$ & $24.3^{+6.6}_{-5.4}$ & $240.8^{+17.2}_{-16.1}$ &	TSH	& 0.82$^{+0.04}_{-0.04}$	& $20.72$ & $0.71$ & $44.35$ 	& 0.6 & 0.1 & $22.00\pm{0.02}^a$ 	& ... \\
23	& J142605.8+353508\space\space\space14:26:05.83\space\space\space35:35:08.7	& 0.4\arcsec	& $246.5^{+17.4}_{-16.3}$ & $154.4^{+13.9}_{-12.8}$ & $92.0^{+11.1}_{-10.0}$ &	TSH	& -0.25$^{+0.07}_{-0.07}$	& $19.39$ & $4.56$ & $16.88$ 	& 0.1 & -0.1 & $20.12\pm{0.00}$ 	& 1.0 \\
24	& J142602.4+353605\space\space\space14:26:02.42\space\space\space35:36:05.6	& 0.6\arcsec	& $81.7^{+10.7}_{-9.6}$ & $61.5^{+9.2}_{-8.1}$ & $20.2^{+6.2}_{-5.0}$ &	TSH	& -0.50$^{+0.12}_{-0.11}$	& $6.50$ & $1.84$ & $3.74$ 	& -0.1 & -0.0 & $21.40\pm{0.01}$ 	& 1.1 \\
25	& J142601.1+354151\space\space\space14:26:01.18\space\space\space35:41:51.1	& 0.7\arcsec	& $284.1^{+18.9}_{-17.8}$ & $179.4^{+15.1}_{-14.0}$ & $104.8^{+12.1}_{-11.0}$ &	TSH	& -0.26$^{+0.06}_{-0.06}$	& $22.87$ & $5.42$ & $19.77$ 	& 0.5 & -0.7 & $20.96\pm{0.01}$ 	& 1.1 \\
26	& J142557.6+353445\space\space\space14:25:57.65\space\space\space35:34:45.9	& 0.4\arcsec	& $65.5^{+9.6}_{-8.5}$ & $40.4^{+7.8}_{-6.6}$ & $25.1^{+6.4}_{-5.2}$ &	TSH	& -0.23$^{+0.14}_{-0.13}$	& $4.92$ & $1.14$ & $4.43$ 	& 0.0 & 0.1 & $19.86\pm{0.00}$ 	& 1.0 \\
27	& J142557.7+353512\space\space\space14:25:57.70\space\space\space35:35:12.4	& 0.9\arcsec	& $13.3^{+5.1}_{-3.9}$ & $ 2.7^{+3.1}_{-1.7}$ & $10.6^{+4.6}_{-3.4}$ &	TH	& 0.60$^{+0.23}_{-0.34}$	& $1.37$ & $0.10$ & $2.58$ 	& 0.6 & -0.5 & $22.65\pm{0.04}$ 	& 1.1 \\
28	& J142556.9+353845\space\space\space14:25:56.94\space\space\space35:38:45.1	& 0.5\arcsec	& $87.6^{+11.0}_{-9.9}$ & $58.3^{+9.0}_{-7.9}$ & $29.4^{+7.0}_{-5.8}$ &	TSH	& -0.33$^{+0.12}_{-0.11}$	& $6.54$ & $1.63$ & $5.12$ 	& 0.2 & -0.3 & $22.42\pm{0.03}$ 	& 1.0 \\
29	& J142556.3+354018\space\space\space14:25:56.34\space\space\space35:40:18.4	& 0.5\arcsec	& $132.0^{+13.1}_{-12.0}$ & $100.1^{+11.5}_{-10.3}$ & $31.8^{+7.2}_{-6.1}$ &	TSH	& -0.51$^{+0.09}_{-0.08}$	& $10.61$ & $3.02$ & $5.98$ 	& -0.0 & -0.2 & $20.81\pm{0.02}$ 	& 1.0 \\
30	& J142555.9+353240\space\space\space14:25:55.95\space\space\space35:32:40.6	& 0.5\arcsec	& $89.4^{+11.1}_{-10.0}$ & $55.2^{+8.9}_{-7.8}$ & $34.2^{+7.4}_{-6.3}$ &	TSH	& -0.23$^{+0.12}_{-0.11}$	& $6.83$ & $1.58$ & $6.13$ 	& 0.1 & 0.2 & $23.06\pm{0.05}$ 	& 1.0 \\
31	& J142555.4+353650\space\space\space14:25:55.40\space\space\space35:36:50.4	& 0.7\arcsec	& $15.3^{+5.5}_{-4.3}$ & $11.9^{+4.8}_{-3.6}$ & $ 3.4^{+3.6}_{-2.3}$ &	TS	& -0.56$^{+0.33}_{-0.25}$	& $1.13$ & $0.33$ & $0.58$ 	& 0.4 & -0.4 & $>$25.7	& ...\\
32	& J142554.1+353237\space\space\space14:25:54.11\space\space\space35:32:37.1	& 0.5\arcsec	& $95.7^{+11.2}_{-10.1}$ & $52.1^{+8.5}_{-7.4}$ & $43.6^{+8.0}_{-6.9}$ &	TSH	& -0.08$^{+0.11}_{-0.11}$	& $7.29$ & $1.49$ & $7.79$ 	& -0.1 & -0.4 & $24.84\pm{0.10}$ 	& 0.9 \\
33	& J142553.8+353826\space\space\space14:25:53.80\space\space\space35:38:26.0	& 0.6\arcsec	& $23.6^{+6.4}_{-5.2}$ & $14.8^{+5.2}_{-4.0}$ & $ 8.8^{+4.5}_{-3.3}$ &	TSH	& -0.25$^{+0.26}_{-0.23}$	& $1.75$ & $0.41$ & $1.53$ 	& -0.5 & -0.5 & $24.44\pm{0.10}$ 	& 0.8 \\
34	& J142553.6+353314\space\space\space14:25:53.68\space\space\space35:33:14.9	& 0.3\arcsec	& $95.8^{+11.3}_{-10.2}$ & $76.7^{+10.1}_{-9.0}$ & $19.1^{+5.8}_{-4.7}$ &	TSH	& -0.60$^{+0.10}_{-0.09}$	& $7.83$ & $2.35$ & $3.67$ 	& -0.0 & -0.2 & $21.41\pm{0.01}$ 	& 1.0 \\
35	& J142552.7+353448\space\space\space14:25:52.72\space\space\space35:34:48.2	& 0.7\arcsec	& $10.0^{+4.6}_{-3.4}$ & $ 7.7^{+4.1}_{-2.9}$ & $ 2.2^{+3.1}_{-1.7}$ &	TS	& -0.54$^{+0.43}_{-0.30}$	& $0.74$ & $0.21$ & $0.39$ 	& -0.3 & 0.2 & $25.21\pm{0.27}$ 	& 1.1 \\
36	& J142552.4+353730\space\space\space14:25:52.45\space\space\space35:37:30.3	& 0.5\arcsec	& $15.2^{+5.5}_{-4.3}$ & $10.6^{+4.6}_{-3.4}$ & $ 4.6^{+3.8}_{-2.5}$ &	TS	& -0.39$^{+0.34}_{-0.28}$	& $1.11$ & $0.29$ & $0.78$ 	& -1.1 & -0.6 & $25.80\pm{0.27}$ 	& 1.3 \\
37	& J142552.2+353823\space\space\space14:25:52.27\space\space\space35:38:23.7	& 0.4\arcsec	& $53.4^{+8.8}_{-7.7}$ & $ 7.4^{+4.1}_{-2.9}$ & $46.0^{+8.2}_{-7.1}$ &	TH	& 0.72$^{+0.10}_{-0.13}$	& $4.10$ & $0.21$ & $8.23$ 	& ... & ... & ...$^b$	& ...\\
38	& J142551.3+353404\space\space\space14:25:51.38\space\space\space35:34:04.4	& 0.4\arcsec	& $37.1^{+7.6}_{-6.5}$ & $31.7^{+7.0}_{-5.8}$ & $ 5.5^{+4.0}_{-2.7}$ &	TS	& -0.70$^{+0.17}_{-0.13}$	& $2.77$ & $0.88$ & $0.96$ 	& 0.1 & -0.1 & $17.38\pm{0.00}$ 	& 1.3 \\
39	& J142551.0+353307\space\space\space14:25:51.00\space\space\space35:33:07.2	& 0.8\arcsec	& $23.2^{+6.5}_{-5.3}$ & $16.2^{+5.5}_{-4.3}$ & $ 7.0^{+4.3}_{-3.1}$ &	TS	& -0.39$^{+0.26}_{-0.23}$	& $1.80$ & $0.47$ & $1.28$ 	& 0.0 & -0.6 & $24.17\pm{0.11}$ 	& 1.0 \\
40	& J142550.8+353033\space\space\space14:25:50.81\space\space\space35:30:33.6	& 1.0\arcsec	& $71.0^{+10.3}_{-9.2}$ & $58.9^{+9.2}_{-8.1}$ & $12.1^{+5.6}_{-4.4}$ &	TSH	& -0.66$^{+0.13}_{-0.11}$	& $5.73$ & $1.78$ & $2.28$ 	& 0.2 & -0.2 & $22.29\pm{0.03}$ 	& 1.0 \\
41	& J142550.6+353743\space\space\space14:25:50.67\space\space\space35:37:43.1	& 0.5\arcsec	& $18.4^{+5.8}_{-4.7}$ & $13.8^{+5.1}_{-3.9}$ & $ 4.6^{+3.8}_{-2.5}$ &	TS	& -0.50$^{+0.29}_{-0.23}$	& $1.34$ & $0.38$ & $0.79$ 	& 0.1 & -0.0 & $23.45\pm{0.06}$ 	& 1.0 \\
42	& J142550.4+353247\space\space\space14:25:50.49\space\space\space35:32:47.9	& 0.8\arcsec	& $26.1^{+6.7}_{-5.5}$ & $20.1^{+5.8}_{-4.7}$ & $ 6.0^{+4.1}_{-2.9}$ &	TS	& -0.53$^{+0.24}_{-0.19}$	& $2.06$ & $0.60$ & $1.12$ 	& 0.0 & -0.1 & $22.93\pm{0.05}$ 	& 1.0 \\
43	& J142549.9+353203\space\space\space14:25:49.90\space\space\space35:32:03.8	& 0.5\arcsec	& $91.9^{+11.1}_{-10.0}$ & $65.1^{+9.4}_{-8.3}$ & $26.8^{+6.7}_{-5.5}$ &	TSH	& -0.41$^{+0.11}_{-0.10}$	& $7.82$ & $2.07$ & $5.37$ 	& 0.1 & -0.1 & $21.69\pm{0.02}$ 	& 1.1 \\
44	& J142549.8+353619\space\space\space14:25:49.81\space\space\space35:36:19.0	& 0.6\arcsec	& $ 9.0^{+4.5}_{-3.3}$ & $ 0.3^{+2.4}_{-0.3}$ & $ 8.7^{+4.3}_{-3.1}$ &	TH	& 0.93$^{+0.06}_{-0.40}$	& $0.69$ & $0.01$ & $1.55$ 	& 0.6 & -0.4 & $23.05\pm{0.07}$ 	& 1.1 \\
45	& J142549.2+353615\space\space\space14:25:49.26\space\space\space35:36:15.3	& 0.6\arcsec	& $ 8.5^{+4.5}_{-3.3}$ & $ 6.1^{+4.0}_{-2.7}$ & $ 2.4^{+3.1}_{-1.7}$ &	TS	& -0.44$^{+0.48}_{-0.35}$	& $0.62$ & $0.17$ & $0.40$ 	& 0.3 & 0.1 & $>$25.7	& ...\\
46	& J142548.5+353507\space\space\space14:25:48.58\space\space\space35:35:07.8	& 0.4\arcsec	& $29.3^{+6.8}_{-5.6}$ & $14.5^{+5.1}_{-3.9}$ & $14.8^{+5.2}_{-4.0}$ &	TSH	& 0.02$^{+0.21}_{-0.22}$	& $2.14$ & $0.40$ & $2.55$ 	& -0.1 & -0.4 & $25.61\pm{0.32}$ 	& 1.4 \\
47	& J142548.2+353041\space\space\space14:25:48.21\space\space\space35:30:41.1	& 0.4\arcsec	& $162.2^{+14.4}_{-13.3}$ & $103.1^{+11.7}_{-10.6}$ & $59.1^{+9.2}_{-8.1}$ &	TSH	& -0.27$^{+0.09}_{-0.08}$	& $12.86$ & $3.06$ & $10.98$ 	& 0.2 & 0.1 & $24.37\pm{0.10}$ 	& 0.7 \\
48	& J142547.6+353043\space\space\space14:25:47.67\space\space\space35:30:43.0	& 0.7\arcsec	& $26.4^{+7.2}_{-6.0}$ & $ 7.6^{+4.5}_{-3.3}$ & $18.8^{+6.2}_{-5.0}$ &	TH	& 0.43$^{+0.21}_{-0.25}$	& $2.14$ & $0.23$ & $3.57$ 	& 0.3 & -0.4 & $22.61\pm{0.04}$ 	& 1.1 \\
49	& J142547.2+353728\space\space\space14:25:47.25\space\space\space35:37:28.0	& 0.9\arcsec	& $ 6.1^{+3.8}_{-2.5}$ & $ 3.2^{+3.1}_{-1.7}$ & $ 2.9^{+3.1}_{-1.7}$ &	T	& -0.04$^{+0.50}_{-0.49}$	& $0.44$ & $0.09$ & $0.49$ 	& ... & ... & $>$25.7	& ...\\
50	& J142546.9+353240\space\space\space14:25:46.95\space\space\space35:32:40.3	& 0.5\arcsec	& $55.4^{+8.8}_{-7.7}$ & $44.3^{+8.0}_{-6.9}$ & $11.1^{+4.6}_{-3.4}$ &	TSH	& -0.59$^{+0.14}_{-0.11}$	& $4.27$ & $1.28$ & $2.01$ 	& -0.0 & -0.4 & $18.41\pm{0.00}$ 	& 2.0 \\
51	& J142546.3+353349\space\space\space14:25:46.33\space\space\space35:33:49.4	& 0.8\arcsec	& $11.8^{+4.9}_{-3.7}$ & $ 2.6^{+3.1}_{-1.7}$ & $ 9.2^{+4.5}_{-3.3}$ &	T	& 0.56$^{+0.26}_{-0.37}$	& $0.90$ & $0.07$ & $1.65$ 	& ... & ... & $>$25.7	& ...\\
52	& J142546.1+353354\space\space\space14:25:46.13\space\space\space35:33:54.9	& 0.4\arcsec	& $17.7^{+5.6}_{-4.4}$ & $10.0^{+4.5}_{-3.3}$ & $ 7.7^{+4.1}_{-2.9}$ &	TSH	& -0.12$^{+0.29}_{-0.28}$	& $1.41$ & $0.30$ & $1.44$ 	& 0.1 & -0.6 & $>$25.7	& ...\\
53	& J142546.0+353826\space\space\space14:25:46.02\space\space\space35:38:26.3	& 0.5\arcsec	& $25.0^{+6.4}_{-5.2}$ & $10.0^{+4.5}_{-3.3}$ & $15.0^{+5.2}_{-4.0}$ &	TSH	& 0.20$^{+0.22}_{-0.24}$	& $1.95$ & $0.29$ & $2.71$ 	& 0.3 & -0.8 & $25.55\pm{0.27}$ 	& 0.9 \\
54	& J142545.8+353849\space\space\space14:25:45.85\space\space\space35:38:49.2	& 0.4\arcsec	& $29.1^{+6.9}_{-5.7}$ & $18.9^{+5.6}_{-4.4}$ & $10.2^{+4.8}_{-3.6}$ &	TSH	& -0.30$^{+0.22}_{-0.20}$	& $2.23$ & $0.54$ & $1.81$ 	& 0.6 & -0.3 & $21.63\pm{0.02}$ 	& 1.4 \\
55	& J142545.7+353228\space\space\space14:25:45.78\space\space\space35:32:28.4	& 0.7\arcsec	& $20.4^{+6.1}_{-4.9}$ & $20.6^{+5.8}_{-4.7}$ & $ 0.0^{+2.8}_{-0.0}$ &	TS	& -1.00$^{+0.21}_{-0.00}$	& $1.54$ & $0.58$ & $<0.49$ 	& 0.1 & -0.2 & $20.28\pm{0.00}$ 	& 1.0 \\
56	& J142545.6+353152\space\space\space14:25:45.69\space\space\space35:31:52.9	& 0.4\arcsec	& $91.9^{+11.3}_{-10.2}$ & $63.9^{+9.4}_{-8.3}$ & $28.0^{+7.1}_{-5.9}$ &	TSH	& -0.38$^{+0.12}_{-0.11}$	& $7.30$ & $1.90$ & $5.24$ 	& -0.1 & -0.0 & $21.96\pm{0.02}$ 	& 1.0 \\
57	& J142545.1+353616\space\space\space14:25:45.17\space\space\space35:36:16.3	& 0.8\arcsec	& $ 6.3^{+3.8}_{-2.5}$ & $ 2.1^{+2.8}_{-1.4}$ & $ 4.2^{+3.3}_{-2.0}$ &	T	& 0.34$^{+0.39}_{-0.52}$	& $0.62$ & $0.08$ & $0.97$ 	& 0.2 & -0.5 & $25.13\pm{0.24}$ 	& 1.2 \\
58	& J142544.7+353954\space\space\space14:25:44.70\space\space\space35:39:54.1	& 0.8\arcsec	& $23.2^{+6.4}_{-5.2}$ & $13.8^{+5.2}_{-4.0}$ & $ 9.3^{+4.5}_{-3.3}$ &	TSH	& -0.19$^{+0.26}_{-0.24}$	& $1.71$ & $0.38$ & $1.61$ 	& 0.0 & -0.2 & $22.55\pm{0.03}$ 	& 1.0 \\
59	& J142541.6+354107\space\space\space14:25:41.66\space\space\space35:41:07.2	& 0.7\arcsec	& $35.3^{+7.8}_{-6.6}$ & $24.4^{+6.5}_{-5.3}$ & $10.8^{+5.1}_{-3.9}$ &	TS	& -0.38$^{+0.21}_{-0.18}$	& $2.79$ & $0.73$ & $2.01$ 	& 0.4 & 0.5 & $24.57\pm{0.19}$ 	& 1.1 \\
60	& J142538.7+353342\space\space\space14:25:38.73\space\space\space35:33:42.1	& 0.4\arcsec	& $14.5^{+5.3}_{-4.2}$ & $ 7.4^{+4.0}_{-2.7}$ & $ 7.2^{+4.3}_{-3.1}$ &	TSH	& -0.01$^{+0.33}_{-0.33}$	& $1.08$ & $0.21$ & $1.26$ 	& -0.0 & -0.3 & $22.06\pm{0.02}$ 	& 1.1 \\
61	& J142538.5+353429\space\space\space14:25:38.51\space\space\space35:34:29.9	& 0.4\arcsec	& $19.4^{+5.9}_{-4.8}$ & $ 4.3^{+3.6}_{-2.3}$ & $15.2^{+5.3}_{-4.2}$ &	TH	& 0.57$^{+0.20}_{-0.27}$	& $1.42$ & $0.12$ & $2.61$ 	& 0.2 & -0.4 & $22.58\pm{0.03}$ 	& 1.0 \\
62	& J142538.2+353438\space\space\space14:25:38.29\space\space\space35:34:38.0	& 0.4\arcsec	& $21.6^{+5.9}_{-4.8}$ & $14.7^{+5.1}_{-3.9}$ & $ 6.9^{+4.0}_{-2.7}$ &	TSH	& -0.36$^{+0.26}_{-0.22}$	& $1.61$ & $0.41$ & $1.21$ 	& 0.2 & -0.2 & $23.56\pm{0.07}$ 	& 1.1 \\
63	& J142537.1+354309\space\space\space14:25:37.15\space\space\space35:43:09.8	& 1.3\arcsec	& $44.6^{+9.1}_{-8.0}$ & $34.9^{+7.7}_{-6.5}$ & $ 9.7^{+5.7}_{-4.5}$ &	TS	& -0.56$^{+0.19}_{-0.16}$	& $3.60$ & $1.06$ & $1.84$ 	& 0.5 & -0.4 & $23.47\pm{0.07}$ 	& 1.1 \\
64	& J142536.9+353209\space\space\space14:25:36.99\space\space\space35:32:09.3	& 0.4\arcsec	& $59.0^{+9.3}_{-8.2}$ & $41.7^{+7.8}_{-6.7}$ & $17.3^{+5.8}_{-4.7}$ &	TSH	& -0.41$^{+0.15}_{-0.13}$	& $4.58$ & $1.21$ & $3.16$ 	& 0.4 & -0.3 & $24.73\pm{0.18}$ 	& 1.1 \\
65	& J142535.9+354107\space\space\space14:25:35.99\space\space\space35:41:07.1	& 0.3\arcsec	& $393.0^{+21.6}_{-20.5}$ & $275.1^{+18.1}_{-17.1}$ & $117.9^{+12.4}_{-11.3}$ &	TSH	& -0.40$^{+0.05}_{-0.05}$	& $30.39$ & $7.99$ & $21.31$ 	& 0.2 & -0.2 & $20.20\pm{0.01}$ 	& 1.0 \\
66	& J142534.5+353551\space\space\space14:25:34.52\space\space\space35:35:51.3	& 0.4\arcsec	& $ 8.2^{+4.1}_{-2.9}$ & $ 6.3^{+3.8}_{-2.5}$ & $ 1.9^{+2.8}_{-1.4}$ &	TS	& -0.54$^{+0.46}_{-0.30}$	& $0.78$ & $0.22$ & $0.41$ 	& 0.1 & 0.1 & $20.68\pm{0.01}$ 	& 1.2 \\
67	& J142533.8+354138\space\space\space14:25:33.80\space\space\space35:41:38.1	& 0.7\arcsec	& $96.8^{+11.6}_{-10.5}$ & $70.9^{+10.0}_{-8.9}$ & $25.8^{+6.7}_{-5.5}$ &	TSH	& -0.46$^{+0.11}_{-0.10}$	& $8.23$ & $2.27$ & $5.15$ 	& -0.1 & -0.3 & $21.26\pm{0.02}$ 	& 1.0 \\
68	& J142533.0+353648\space\space\space14:25:33.09\space\space\space35:36:48.5	& 0.5\arcsec	& $18.2^{+5.6}_{-4.4}$ & $13.2^{+4.9}_{-3.7}$ & $ 5.0^{+3.6}_{-2.3}$ &	TSH	& -0.45$^{+0.28}_{-0.22}$	& $2.20$ & $0.60$ & $1.40$ 	& 0.2 & -0.0 & $25.00\pm{0.21}$ 	& 0.8 \\
69	& J142531.1+353921\space\space\space14:25:31.17\space\space\space35:39:21.6	& 0.6\arcsec	& $36.4^{+7.6}_{-6.5}$ & $ 4.9^{+3.8}_{-2.5}$ & $31.5^{+7.1}_{-5.9}$ &	TH	& 0.73$^{+0.12}_{-0.17}$	& $2.82$ & $0.14$ & $5.68$ 	& ... & ... & $>$25.7	& ...\\
70	& J142530.6+353420\space\space\space14:25:30.63\space\space\space35:34:20.3	& 0.3\arcsec	& $48.3^{+8.3}_{-7.2}$ & $34.5^{+7.2}_{-6.0}$ & $13.8^{+5.1}_{-3.9}$ &	TSH	& -0.42$^{+0.16}_{-0.14}$	& $3.53$ & $0.94$ & $2.36$ 	& ... & ... & $>$25.7	& ...\\
71	& J142529.9+353640\space\space\space14:25:29.91\space\space\space35:36:40.0	& 0.4\arcsec	& $16.9^{+5.5}_{-4.3}$ & $12.9^{+4.9}_{-3.7}$ & $ 4.0^{+3.3}_{-2.0}$ &	TS	& -0.53$^{+0.29}_{-0.22}$	& $1.29$ & $0.37$ & $0.70$ 	& 0.1 & -0.5 & $25.80\pm{0.29}$ 	& 0.8 \\
72	& J142529.2+353248\space\space\space14:25:29.22\space\space\space35:32:48.4	& 0.4\arcsec	& $40.4^{+7.8}_{-6.7}$ & $29.8^{+6.9}_{-5.7}$ & $10.6^{+4.6}_{-3.4}$ &	TSH	& -0.47$^{+0.18}_{-0.15}$	& $3.04$ & $0.84$ & $1.87$ 	& ... & ... & $>$25.7	& ...\\
73	& J142526.6+353140\space\space\space14:25:26.68\space\space\space35:31:40.8	& 0.8\arcsec	& $42.3^{+8.2}_{-7.0}$ & $24.1^{+6.4}_{-5.2}$ & $18.1^{+5.8}_{-4.7}$ &	TSH	& -0.13$^{+0.19}_{-0.18}$	& $3.29$ & $0.70$ & $3.32$ 	& ... & ... & $>$25.7	& ...\\
74	& J142525.7+353841\space\space\space14:25:25.72\space\space\space35:38:41.5	& 0.4\arcsec	& $62.8^{+9.4}_{-8.3}$ & $ 8.7^{+4.3}_{-3.1}$ & $54.1^{+8.8}_{-7.6}$ &	TSH	& 0.72$^{+0.09}_{-0.12}$	& $4.66$ & $0.24$ & $9.33$ 	& 0.6 & -0.5 & $24.79\pm{0.25}$ 	& 1.9 \\
75	& J142524.7+353037\space\space\space14:25:24.70\space\space\space35:30:37.1	& 0.5\arcsec	& $151.2^{+14.0}_{-12.9}$ & $117.3^{+12.3}_{-11.2}$ & $33.9^{+7.5}_{-6.4}$ &	TSH	& -0.54$^{+0.08}_{-0.07}$	& $12.30$ & $3.57$ & $6.49$ 	& 0.7 & 0.1 & $23.46\pm{0.07}$ 	& 1.1 \\
76	& J142523.6+352824\space\space\space14:25:23.69\space\space\space35:28:24.5	& 0.7\arcsec	& $301.7^{+19.1}_{-18.0}$ & $58.9^{+9.1}_{-8.0}$ & $242.8^{+17.2}_{-16.1}$ &	TSH	& 0.62$^{+0.05}_{-0.05}$	& $85.17$ & $6.21$ & $161.16$ 	& 0.2 & -0.1 & $18.51\pm{0.00}$ 	& 1.1 \\
77	& J142521.8+353407\space\space\space14:25:21.87\space\space\space35:34:07.8	& 0.5\arcsec	& $35.3^{+7.5}_{-6.4}$ & $20.1^{+5.8}_{-4.7}$ & $15.2^{+5.5}_{-4.3}$ &	TSH	& -0.14$^{+0.20}_{-0.20}$	& $2.78$ & $0.59$ & $2.78$ 	& 0.1 & -0.4 & $25.61\pm{0.29}$ 	& 1.0 \\
78	& J142521.3+353029\space\space\space14:25:21.38\space\space\space35:30:29.5	& 0.9\arcsec	& $51.6^{+9.2}_{-8.1}$ & $15.1^{+5.7}_{-4.5}$ & $36.5^{+7.8}_{-6.7}$ &	TSH	& 0.42$^{+0.15}_{-0.17}$	& $4.26$ & $0.47$ & $7.06$ 	& -0.4 & -0.2 & $21.52\pm{0.01}$ 	& 1.1 \\
79	& J142517.7+353822\space\space\space14:25:17.70\space\space\space35:38:22.2	& 0.7\arcsec	& $37.3^{+7.6}_{-6.5}$ & $22.2^{+6.1}_{-4.9}$ & $15.1^{+5.3}_{-4.2}$ &	TSH	& -0.19$^{+0.19}_{-0.18}$	& $2.83$ & $0.63$ & $2.67$ 	& 0.5 & -0.3 & $19.19\pm{0.00}$ 	& 1.8 \\
80	& J142514.3+353918\space\space\space14:25:14.37\space\space\space35:39:18.2	& 0.6\arcsec	& $118.5^{+12.6}_{-11.5}$ & $71.7^{+9.9}_{-8.8}$ & $46.8^{+8.5}_{-7.4}$ &	TSH	& -0.21$^{+0.10}_{-0.10}$	& $9.17$ & $2.09$ & $8.45$ 	& 0.5 & -0.9 & $23.71\pm{0.09}$ 	& 1.2 \\
81	& J142514.3+353618\space\space\space14:25:14.34\space\space\space35:36:18.8	& 0.4\arcsec	& $102.9^{+11.7}_{-10.6}$ & $82.2^{+10.4}_{-9.3}$ & $20.7^{+6.2}_{-5.0}$ &	TSH	& -0.59$^{+0.10}_{-0.09}$	& $7.83$ & $2.34$ & $3.68$ 	& 0.1 & -0.2 & $19.86\pm{0.00}$ 	& 1.2 \\
82	& J142513.1+353323\space\space\space14:25:13.18\space\space\space35:33:23.4	& 0.6\arcsec	& $98.5^{+11.8}_{-10.7}$ & $73.9^{+10.2}_{-9.1}$ & $24.6^{+6.7}_{-5.5}$ &	TSH	& -0.50$^{+0.11}_{-0.10}$	& $8.18$ & $2.30$ & $4.77$ 	& 0.4 & -0.3 & $18.51\pm{0.00}$ 	& 1.4 \\
83	& J142512.0+353124\space\space\space14:25:12.00\space\space\space35:31:24.3	& 0.7\arcsec	& $214.7^{+16.5}_{-15.5}$ & $165.1^{+14.5}_{-13.4}$ & $49.5^{+8.8}_{-7.7}$ &	TSH	& -0.53$^{+0.07}_{-0.06}$	& $17.70$ & $5.09$ & $9.57$ 	& 0.6 & 0.1 & $20.86\pm{0.01}$ 	& 1.0 \\
84	& J142511.3+353857\space\space\space14:25:11.32\space\space\space35:38:57.8	& 0.6\arcsec	& $121.6^{+12.9}_{-11.8}$ & $89.7^{+11.0}_{-9.9}$ & $31.9^{+7.6}_{-6.5}$ &	TSH	& -0.47$^{+0.10}_{-0.09}$	& $9.49$ & $2.63$ & $5.82$ 	& -0.2 & 0.2 & $24.88\pm{0.26}$ 	& 1.5 \\
85	& J142509.3+354356\space\space\space14:25:09.39\space\space\space35:43:56.1	& 1.1\arcsec	& $349.8^{+21.1}_{-20.0}$ & $267.4^{+18.3}_{-17.2}$ & $82.5^{+11.4}_{-10.3}$ &	TSH	& -0.52$^{+0.05}_{-0.05}$	& $41.92$ & $12.05$ & $23.30$ 	& 0.2 & -0.6 & $20.13\pm{0.01}$ 	& 1.2 \\
86	& J142504.5+354107\space\space\space14:25:04.58\space\space\space35:41:07.9	& 0.6\arcsec	& $473.1^{+23.9}_{-22.8}$ & $338.3^{+20.2}_{-19.1}$ & $134.9^{+13.5}_{-12.4}$ &	TSH	& -0.42$^{+0.05}_{-0.04}$	& $41.21$ & $11.08$ & $27.65$ 	& 0.6 & 0.3 & $21.39\pm{0.01}$ 	& 1.1 \\
87	& J142457.1+353518\space\space\space14:24:57.16\space\space\space35:35:18.8	& 0.9\arcsec	& $257.4^{+18.2}_{-17.1}$ & $197.9^{+15.8}_{-14.7}$ & $59.6^{+9.9}_{-8.8}$ &	TSH	& -0.53$^{+0.06}_{-0.06}$	& $24.83$ & $7.15$ & $13.52$ 	& 0.5 & -0.7 & $20.00\pm{0.00}$ 	& 1.0 \\
88	& J142618.8+354218\space\space\space14:26:18.84\space\space\space35:42:18.4	& 1.5\arcsec	& $227.6^{+17.8}_{-16.7}$ & $156.9^{+14.4}_{-13.3}$ & $70.7^{+11.1}_{-10.0}$ &	TSH	& -0.37$^{+0.07}_{-0.07}$	& $26.93$ & $6.97$ & $19.76$ 	& 0.9 & 0.1 & $20.89\pm{0.01}$ 	& 1.1 \\
89	& J142618.6+353754\space\space\space14:26:18.60\space\space\space35:37:54.2	& 2.2\arcsec	& $39.8^{+9.4}_{-8.3}$ & $ 9.1^{+5.2}_{-4.0}$ & $30.7^{+8.3}_{-7.2}$ &	TH	& 0.55$^{+0.17}_{-0.21}$	& $3.24$ & $0.28$ & $5.86$ 	& -0.3 & -0.7 & $21.77\pm{0.02}$ 	& 1.2 \\
90	& J142611.9+354230\space\space\space14:26:11.92\space\space\space35:42:30.5	& 1.7\arcsec	& $80.0^{+12.1}_{-11.0}$ & $62.4^{+9.9}_{-8.8}$ & $17.6^{+7.8}_{-6.6}$ &	TS	& -0.56$^{+0.14}_{-0.13}$	& $9.28$ & $2.72$ & $4.80$ 	& 0.6 & -0.0 & $21.40\pm{0.01}$ 	& 1.0 \\
91	& J142610.8+354202\space\space\space14:26:10.89\space\space\space35:42:02.5	& 1.5\arcsec	& $151.9^{+14.8}_{-13.7}$ & $45.4^{+8.8}_{-7.6}$ & $106.6^{+12.4}_{-11.3}$ &	TSH	& 0.41$^{+0.08}_{-0.09}$	& $17.47$ & $1.96$ & $28.81$ 	& 1.1 & -0.2 & $20.88\pm{0.02}$ 	& 1.3 \\
92	& J142606.7+353151\space\space\space14:26:06.72\space\space\space35:31:51.3	& 1.5\arcsec	& $50.3^{+9.3}_{-8.2}$ & $36.2^{+7.8}_{-6.6}$ & $14.0^{+5.9}_{-4.8}$ &	TS	& -0.44$^{+0.18}_{-0.16}$	& $4.00$ & $1.08$ & $2.63$ 	& -0.3 & 0.3 & $24.70\pm{0.21}$ 	& 1.1 \\
93	& J142604.7+353015\space\space\space14:26:04.78\space\space\space35:30:15.3	& 1.7\arcsec	& $75.8^{+11.0}_{-9.9}$ & $34.6^{+7.6}_{-6.5}$ & $41.3^{+8.5}_{-7.4}$ &	TSH	& 0.10$^{+0.14}_{-0.14}$	& $6.90$ & $1.18$ & $8.84$ 	& 1.1 & -0.2 & $21.15\pm{0.01}$ 	& 1.2 \\
94	& J142603.7+353246\space\space\space14:26:03.73\space\space\space35:32:46.2	& 1.6\arcsec	& $15.0^{+6.2}_{-5.0}$ & $ 6.0^{+4.1}_{-2.9}$ & $ 9.0^{+5.2}_{-4.0}$ &	T	& 0.21$^{+0.34}_{-0.38}$	& $1.24$ & $0.19$ & $1.75$ 	& -0.4 & 0.1 & $21.83\pm{0.02}$ 	& 1.3 \\
95	& J142600.2+353442\space\space\space14:26:00.24\space\space\space35:34:42.4	& 0.8\arcsec	& $17.3^{+5.7}_{-4.5}$ & $ 0.0^{+1.9}_{-0.0}$ & $18.3^{+5.7}_{-4.5}$ &	TH	& 1.00$^{+0.00}_{-0.18}$	& $1.34$ & $<0.06$ & $3.31$ 	& 0.3 & -0.6 & $24.33\pm{0.13}$ 	& 1.1 \\
96	& J142558.1+353216\space\space\space14:25:58.13\space\space\space35:32:16.1	& 1.5\arcsec	& $19.6^{+6.3}_{-5.1}$ & $13.1^{+5.2}_{-4.0}$ & $ 6.5^{+4.3}_{-3.1}$ &	TS	& -0.33$^{+0.30}_{-0.26}$	& $1.52$ & $0.38$ & $1.18$ 	& ... & ... & $>$25.7	& ...\\
97	& J142552.4+352724\space\space\space14:25:52.47\space\space\space35:27:24.1	& 1.8\arcsec	& $88.1^{+11.9}_{-10.8}$ & $50.6^{+8.8}_{-7.7}$ & $37.5^{+8.6}_{-7.5}$ &	TSH	& -0.14$^{+0.13}_{-0.13}$	& $10.55$ & $2.27$ & $10.58$ 	& -0.6 & -0.5 & $23.27\pm{0.08}$ 	& 1.0 \\
98	& J142551.4+353433\space\space\space14:25:51.43\space\space\space35:34:33.1	& 0.7\arcsec	& $ 9.6^{+4.5}_{-3.3}$ & $ 0.0^{+1.9}_{-0.0}$ & $ 9.8^{+4.5}_{-3.3}$ &	TH	& 1.00$^{+0.00}_{-0.31}$	& $0.71$ & $<0.05$ & $1.72$ 	& 0.1 & -0.1 & $24.57\pm{0.18}$ 	& 1.3 \\
99	& J142551.1+353015\space\space\space14:25:51.16\space\space\space35:30:15.7	& 1.5\arcsec	& $20.6^{+6.9}_{-5.7}$ & $16.1^{+5.7}_{-4.5}$ & $ 4.5^{+4.6}_{-3.4}$ &	TS	& -0.56$^{+0.32}_{-0.26}$	& $1.62$ & $0.48$ & $0.83$ 	& 0.1 & 0.2 & $25.68\pm{0.28}$ 	& 0.9 \\
100	& J142547.4+352720\space\space\space14:25:47.43\space\space\space35:27:20.2	& 1.4\arcsec	& $179.7^{+15.9}_{-14.8}$ & $127.4^{+13.1}_{-12.0}$ & $52.3^{+9.8}_{-8.7}$ &	TSH	& -0.41$^{+0.08}_{-0.08}$	& $15.14$ & $4.03$ & $10.38$ 	& 0.4 & -0.3 & $19.87\pm{0.00}$ 	& 1.0 \\
101	& J142545.5+353003\space\space\space14:25:45.53\space\space\space35:30:03.6	& 1.3\arcsec	& $22.8^{+6.8}_{-5.6}$ & $11.7^{+4.9}_{-3.7}$ & $11.0^{+5.3}_{-4.2}$ &	TS	& -0.03$^{+0.27}_{-0.28}$	& $1.85$ & $0.36$ & $2.11$ 	& 0.7 & 0.6 & $23.73\pm{0.09}$ 	& 1.0 \\
102	& J142544.7+354046\space\space\space14:25:44.77\space\space\space35:40:46.6	& 1.1\arcsec	& $20.4^{+6.1}_{-4.9}$ & $ 6.4^{+4.0}_{-2.7}$ & $14.0^{+5.2}_{-4.0}$ &	TH	& 0.38$^{+0.23}_{-0.28}$	& $1.61$ & $0.19$ & $2.61$ 	& 0.1 & -0.7 & $20.09\pm{0.01}$ 	& 1.2 \\
103	& J142544.2+354018\space\space\space14:25:44.20\space\space\space35:40:18.4	& 0.9\arcsec	& $22.7^{+6.2}_{-5.0}$ & $ 1.6^{+2.8}_{-1.4}$ & $21.1^{+5.9}_{-4.8}$ &	TH	& 0.86$^{+0.10}_{-0.20}$	& $1.72$ & $0.05$ & $3.74$ 	& ... & ... & $>$25.7	& ...\\
104	& J142541.8+354249\space\space\space14:25:41.80\space\space\space35:42:49.7	& 1.3\arcsec	& $42.3^{+8.4}_{-7.3}$ & $24.1^{+6.4}_{-5.2}$ & $18.2^{+6.2}_{-5.0}$ &	TSH	& -0.13$^{+0.19}_{-0.18}$	& $3.64$ & $0.78$ & $3.68$ 	& -0.9 & 0.6 & $21.49\pm{0.02}$ 	& 1.2 \\
105	& J142539.6+353942\space\space\space14:25:39.65\space\space\space35:39:42.6	& 0.8\arcsec	& $18.3^{+5.7}_{-4.5}$ & $11.3^{+4.6}_{-3.4}$ & $ 7.0^{+4.1}_{-2.9}$ &	TS	& -0.24$^{+0.29}_{-0.27}$	& $1.37$ & $0.32$ & $1.21$ 	& 0.4 & -0.5 & $25.49\pm{0.30}$ 	& 1.1 \\
106	& J142536.8+353147\space\space\space14:25:36.81\space\space\space35:31:47.2	& 1.0\arcsec	& $11.5^{+5.1}_{-3.9}$ & $ 2.1^{+3.1}_{-1.7}$ & $ 9.3^{+4.6}_{-3.4}$ &	TH	& 0.63$^{+0.25}_{-0.39}$	& $0.91$ & $0.06$ & $1.73$ 	& 0.2 & -0.2 & $24.88\pm{0.21}$ 	& 2.0 \\
107	& J142533.3+353130\space\space\space14:25:33.33\space\space\space35:31:30.6	& 0.8\arcsec	& $20.7^{+6.4}_{-5.2}$ & $ 5.3^{+3.8}_{-2.5}$ & $15.4^{+5.7}_{-4.5}$ &	TH	& 0.49$^{+0.22}_{-0.28}$	& $1.59$ & $0.15$ & $2.78$ 	& 0.5 & -0.2 & $22.50\pm{0.03}$ 	& 1.0 \\
108	& J142532.6+352838\space\space\space14:25:32.66\space\space\space35:28:38.8	& 1.4\arcsec	& $39.5^{+9.0}_{-7.9}$ & $32.9^{+7.7}_{-6.5}$ & $ 6.6^{+5.5}_{-4.3}$ &	TS	& -0.66$^{+0.21}_{-0.17}$	& $3.21$ & $1.00$ & $1.26$ 	& 0.9 & 0.9 & $24.35\pm{0.17}$ 	& 1.4 \\
109	& J142531.4+354452\space\space\space14:25:31.41\space\space\space35:44:52.5	& 2.1\arcsec	& $62.7^{+10.8}_{-9.7}$ & $20.7^{+6.5}_{-5.3}$ & $42.0^{+9.2}_{-8.1}$ &	TSH	& 0.35$^{+0.15}_{-0.16}$	& $5.35$ & $0.66$ & $8.49$ 	& -0.0 & 1.4 & $20.02\pm{0.00}$ 	& 1.2 \\
110	& J142530.8+353208\space\space\space14:25:30.82\space\space\space35:32:08.6	& 1.1\arcsec	& $10.3^{+4.5}_{-3.3}$ & $ 5.0^{+3.6}_{-2.3}$ & $ 5.3^{+3.6}_{-2.3}$ &	T	& 0.03$^{+0.37}_{-0.38}$	& $1.31$ & $0.24$ & $1.58$ 	& -0.1 & -0.4 & $20.91\pm{0.01}$ 	& 1.2 \\
111	& J142528.2+353958\space\space\space14:25:28.24\space\space\space35:39:58.4	& 0.9\arcsec	& $21.1^{+6.2}_{-5.0}$ & $ 3.8^{+3.6}_{-2.3}$ & $17.3^{+5.6}_{-4.4}$ &	TH	& 0.64$^{+0.19}_{-0.26}$	& $1.58$ & $0.11$ & $3.01$ 	& 0.5 & -0.3 & $22.81\pm{0.04}$ 	& 1.0 \\
112	& J142527.5+354012\space\space\space14:25:27.59\space\space\space35:40:12.1	& 0.8\arcsec	& $24.6^{+6.8}_{-5.6}$ & $11.2^{+4.8}_{-3.6}$ & $13.5^{+5.5}_{-4.3}$ &	TSH	& 0.10$^{+0.25}_{-0.26}$	& $1.89$ & $0.32$ & $2.41$ 	& ... & ... & $>$25.7	& ...\\
113	& J142527.4+353101\space\space\space14:25:27.49\space\space\space35:31:01.0	& 1.2\arcsec	& $12.7^{+5.6}_{-4.4}$ & $ 8.6^{+4.6}_{-3.4}$ & $ 4.1^{+4.0}_{-2.7}$ &	TS	& -0.34$^{+0.41}_{-0.35}$	& $1.06$ & $0.27$ & $0.81$ 	& 0.4 & -0.1 & $20.38\pm{0.02}$ 	& 1.3 \\
114	& J142526.6+353327\space\space\space14:25:26.66\space\space\space35:33:27.2	& 0.7\arcsec	& $10.8^{+4.9}_{-3.7}$ & $ 8.3^{+4.3}_{-3.1}$ & $ 2.5^{+3.3}_{-2.0}$ &	TS	& -0.53$^{+0.42}_{-0.31}$	& $0.84$ & $0.24$ & $0.46$ 	& -1.0 & -0.1 & $24.91\pm{0.15}$ 	& 1.2 \\
115	& J142524.2+352928\space\space\space14:25:24.22\space\space\space35:29:28.6	& 1.4\arcsec	& $44.8^{+9.0}_{-7.9}$ & $25.9^{+6.7}_{-5.5}$ & $18.9^{+6.7}_{-5.5}$ &	TSH	& -0.14$^{+0.19}_{-0.19}$	& $4.00$ & $0.86$ & $4.01$ 	& ... & ... & $>$25.7	& ...\\
116	& J142523.8+354122\space\space\space14:25:23.80\space\space\space35:41:22.4	& 0.8\arcsec	& $52.7^{+9.2}_{-8.1}$ & $41.9^{+8.0}_{-6.9}$ & $10.8^{+5.3}_{-4.2}$ &	TS	& -0.59$^{+0.16}_{-0.13}$	& $4.26$ & $1.27$ & $2.04$ 	& 0.0 & 0.2 & $22.98\pm{0.04}$ 	& 1.0 \\
117	& J142523.6+353216\space\space\space14:25:23.69\space\space\space35:32:16.7	& 1.1\arcsec	& $26.5^{+6.7}_{-5.5}$ & $11.4^{+4.9}_{-3.7}$ & $15.0^{+5.2}_{-4.0}$ &	TSH	& 0.14$^{+0.23}_{-0.24}$	& $2.06$ & $0.33$ & $2.74$ 	& 0.6 & -0.5 & $25.34\pm{0.27}$ 	& 0.8 \\
118	& J142523.0+354313\space\space\space14:25:23.00\space\space\space35:43:13.4	& 1.1\arcsec	& $104.3^{+12.6}_{-11.5}$ & $66.8^{+9.8}_{-8.7}$ & $37.5^{+8.5}_{-7.4}$ &	TSH	& -0.28$^{+0.12}_{-0.11}$	& $8.28$ & $1.99$ & $7.00$ 	& 0.3 & 0.8 & $21.42\pm{0.02}$ 	& 1.0 \\
119	& J142522.8+353116\space\space\space14:25:22.82\space\space\space35:31:16.8	& 1.0\arcsec	& $28.9^{+7.1}_{-5.9}$ & $ 4.6^{+3.8}_{-2.5}$ & $24.3^{+6.5}_{-5.3}$ &	TH	& 0.68$^{+0.15}_{-0.21}$	& $2.44$ & $0.15$ & $4.79$ 	& 0.7 & -0.2 & $23.41\pm{0.07}$ 	& 1.1 \\
120	& J142522.4+353517\space\space\space14:25:22.40\space\space\space35:35:17.2	& 0.6\arcsec	& $10.6^{+4.8}_{-3.6}$ & $ 3.2^{+3.3}_{-2.0}$ & $ 7.4^{+4.1}_{-2.9}$ &	T	& 0.40$^{+0.32}_{-0.41}$	& $0.81$ & $0.09$ & $1.32$ 	& ... & ... & $>$25.7	& ...\\
121	& J142521.1+353340\space\space\space14:25:21.10\space\space\space35:33:40.0	& 0.7\arcsec	& $16.3^{+5.6}_{-4.4}$ & $14.1^{+5.2}_{-4.0}$ & $ 2.2^{+3.1}_{-1.7}$ &	TS	& -0.73$^{+0.29}_{-0.18}$	& $1.59$ & $0.51$ & $0.50$ 	& 0.0 & -0.6 & $25.03\pm{0.21}$ 	& 1.1 \\
122	& J142517.6+353453\space\space\space14:25:17.61\space\space\space35:34:53.0	& 0.5\arcsec	& $19.1^{+5.9}_{-4.8}$ & $11.5^{+4.8}_{-3.6}$ & $ 7.6^{+4.3}_{-3.1}$ &	TS	& -0.20$^{+0.29}_{-0.27}$	& $1.44$ & $0.33$ & $1.34$ 	& 0.7 & 0.3 & $24.03\pm{0.11}$ 	& 1.0 \\
        &
        &                       &                 &                           &
                &                                       &       &        &
  &             & -0.2 & -1.3 & $25.08\pm{0.09}$ & 1.3 \\
123	& J142516.6+354129\space\space\space14:25:16.67\space\space\space35:41:29.6	& 1.3\arcsec	& $45.8^{+8.6}_{-7.5}$ & $33.1^{+7.2}_{-6.1}$ & $12.7^{+5.5}_{-4.3}$ &	TS	& -0.44$^{+0.18}_{-0.16}$	& $3.60$ & $0.98$ & $2.34$ 	& 0.2 & 0.1 & $22.49\pm{0.04}$ 	& 1.2 \\
124	& J142516.4+352940\space\space\space14:25:16.43\space\space\space35:29:40.7	& 1.2\arcsec	& $70.9^{+10.7}_{-9.6}$ & $47.5^{+8.5}_{-7.4}$ & $23.4^{+7.2}_{-6.1}$ &	TSH	& -0.33$^{+0.15}_{-0.14}$	& $6.16$ & $1.54$ & $4.78$ 	& 0.2 & 0.8 & $21.31\pm{0.01}$ 	& 1.2 \\
125	& J142516.0+353142\space\space\space14:25:16.00\space\space\space35:31:42.3	& 1.1\arcsec	& $11.6^{+5.7}_{-4.5}$ & $ 7.1^{+4.3}_{-3.1}$ & $ 4.5^{+4.5}_{-3.3}$ &	T	& -0.22$^{+0.45}_{-0.43}$	& $0.97$ & $0.22$ & $0.89$ 	& -0.1 & -0.8 & $24.29\pm{0.16}$ 	& 1.0 \\
126	& J142514.9+354004\space\space\space14:25:14.90\space\space\space35:40:04.1	& 0.9\arcsec	& $48.1^{+8.9}_{-7.8}$ & $31.3^{+7.2}_{-6.0}$ & $16.8^{+6.1}_{-4.9}$ &	TSH	& -0.30$^{+0.18}_{-0.16}$	& $3.74$ & $0.92$ & $3.06$ 	& 1.0 & 0.2 & $22.91\pm{0.05}$ 	& 1.2 \\
127	& J142510.9+353602\space\space\space14:25:10.94\space\space\space35:36:02.8	& 1.0\arcsec	& $39.7^{+8.0}_{-6.9}$ & $16.4^{+5.5}_{-4.3}$ & $23.2^{+6.5}_{-5.3}$ &	TSH	& 0.18$^{+0.18}_{-0.19}$	& $3.08$ & $0.48$ & $4.21$ 	& -0.9 & 0.3 & $22.67\pm{0.05}$ 	& 1.0 \\
128	& J142510.1+353411\space\space\space14:25:10.19\space\space\space35:34:11.4	& 0.9\arcsec	& $50.4^{+8.9}_{-7.8}$ & $31.2^{+7.1}_{-5.9}$ & $19.2^{+6.2}_{-5.0}$ &	TSH	& -0.23$^{+0.17}_{-0.16}$	& $3.94$ & $0.91$ & $3.51$ 	& -0.6 & 0.1 & $21.87\pm{0.02}$ 	& 1.2 \\
129	& J142509.5+354244\space\space\space14:25:09.51\space\space\space35:42:44.9	& 2.1\arcsec	& $71.2^{+11.5}_{-10.3}$ & $15.2^{+6.3}_{-5.1}$ & $56.0^{+10.1}_{-9.0}$ &	TH	& 0.58$^{+0.12}_{-0.14}$	& $7.45$ & $0.60$ & $13.80$ 	& -0.8 & 2.0 & $20.77\pm{0.01}$ 	& 1.4 \\
130	& J142509.5+353747\space\space\space14:25:09.50\space\space\space35:37:47.5	& 0.6\arcsec	& $110.9^{+12.4}_{-11.3}$ & $32.6^{+7.2}_{-6.1}$ & $78.3^{+10.6}_{-9.5}$ &	TSH	& 0.42$^{+0.09}_{-0.10}$	& $8.62$ & $0.95$ & $14.24$ 	& 0.5 & -0.1 & $21.93\pm{0.01}$ 	& 1.2 \\
131	& J142509.2+353527\space\space\space14:25:09.23\space\space\space35:35:27.5	& 0.8\arcsec	& $71.0^{+10.1}_{-9.0}$ & $ 9.2^{+4.6}_{-3.4}$ & $61.8^{+9.4}_{-8.3}$ &	TSH	& 0.74$^{+0.09}_{-0.11}$	& $5.53$ & $0.27$ & $11.24$ 	& -0.3 & -0.0 & $20.85\pm{0.01}$ 	& 1.0 \\
132	& J142505.1+354054\space\space\space14:25:05.16\space\space\space35:40:54.9	& 0.8\arcsec	& $205.0^{+16.4}_{-15.3}$ & $131.2^{+13.2}_{-12.1}$ & $73.8^{+10.4}_{-9.3}$ &	TSH	& -0.27$^{+0.08}_{-0.07}$	& $17.72$ & $4.27$ & $15.01$ 	& -0.5 & 0.6 & $22.59\pm{0.04}$ 	& 1.0 \\
133	& J142504.4+353514\space\space\space14:25:04.47\space\space\space35:35:14.6	& 1.5\arcsec	& $39.9^{+8.6}_{-7.5}$ & $26.7^{+6.8}_{-5.6}$ & $13.2^{+6.1}_{-4.9}$ &	TS	& -0.33$^{+0.21}_{-0.19}$	& $3.17$ & $0.80$ & $2.45$ 	& 0.3 & -0.7 & $23.16\pm{0.06}$ 	& 1.0 \\
134	& J142503.9+353345\space\space\space14:25:03.90\space\space\space35:33:45.7	& 1.5\arcsec	& $42.4^{+8.7}_{-7.6}$ & $13.5^{+5.5}_{-4.3}$ & $28.9^{+7.3}_{-6.2}$ &	TSH	& 0.37$^{+0.17}_{-0.19}$	& $4.40$ & $0.52$ & $7.04$ 	& -0.9 & -0.2 & $20.29\pm{0.00}$ 	& 1.0 \\
135	& J142503.1+353421\space\space\space14:25:03.16\space\space\space35:34:21.2	& 1.2\arcsec	& $60.3^{+10.1}_{-9.0}$ & $13.5^{+5.8}_{-4.7}$ & $46.8^{+8.8}_{-7.7}$ &	TSH	& 0.56$^{+0.13}_{-0.15}$	& $4.85$ & $0.41$ & $8.82$ 	& 1.1 & 0.5 & $>$25.7	& ...\\
136	& J142502.9+353512\space\space\space14:25:02.94\space\space\space35:35:12.8	& 1.0\arcsec	& $112.1^{+12.7}_{-11.6}$ & $69.9^{+10.0}_{-8.9}$ & $42.2^{+8.5}_{-7.3}$ &	TSH	& -0.24$^{+0.11}_{-0.11}$	& $9.08$ & $2.12$ & $8.00$ 	& -0.6 & -0.2 & $25.11\pm{0.31}$ 	& 1.1 \\
137	& J142501.2+353622\space\space\space14:25:01.20\space\space\space35:36:22.0	& 1.2\arcsec	& $93.5^{+11.9}_{-10.8}$ & $66.5^{+9.9}_{-8.8}$ & $27.1^{+7.4}_{-6.3}$ &	TSH	& -0.42$^{+0.12}_{-0.11}$	& $7.82$ & $2.08$ & $5.31$ 	& 0.5 & 0.0 & $23.12\pm{0.05}$ 	& 1.0 \\
138	& J142457.3+353627\space\space\space14:24:57.33\space\space\space35:36:27.5	& 1.2\arcsec	& $93.5^{+12.1}_{-11.0}$ & $45.3^{+8.6}_{-7.5}$ & $48.1^{+9.2}_{-8.1}$ &	TSH	& 0.04$^{+0.12}_{-0.13}$	& $7.63$ & $1.39$ & $9.24$ 	& 0.5 & 0.2 & $>$25.7	& ...\\
139	& J142454.8+353431\space\space\space14:24:54.80\space\space\space35:34:31.9	& 0.9\arcsec	& $250.3^{+18.3}_{-17.2}$ & $157.6^{+14.4}_{-13.3}$ & $92.7^{+12.0}_{-10.9}$ &	TSH	& -0.25$^{+0.07}_{-0.07}$	& $22.73$ & $5.36$ & $19.83$ 	& 0.2 & -0.7 & $20.50\pm{0.01}$ 	& 1.1 \\
140	& J142616.9+353729\space\space\space14:26:16.94\space\space\space35:37:29.4	& 1.9\arcsec	& $71.1^{+10.9}_{-9.8}$ & $23.4^{+6.9}_{-5.7}$ & $47.6^{+9.0}_{-7.9}$ &	TSH	& 0.35$^{+0.13}_{-0.15}$	& $5.91$ & $0.73$ & $9.29$ 	& 0.9 & 0.7 & $23.11\pm{0.06}$ 	& 1.5 \\
141	& J142613.9+353445\space\space\space14:26:13.97\space\space\space35:34:45.5	& 1.6\arcsec	& $35.4^{+8.6}_{-7.5}$ & $26.6^{+6.9}_{-5.7}$ & $ 8.8^{+5.9}_{-4.8}$ &	TS	& -0.50$^{+0.24}_{-0.21}$	& $2.81$ & $0.79$ & $1.64$ 	& ... & ... & $>$25.7	& ...\\
142	& J142601.7+354037\space\space\space14:26:01.72\space\space\space35:40:37.5	& 1.3\arcsec	& $19.4^{+6.6}_{-5.4}$ & $12.4^{+5.1}_{-3.9}$ & $ 7.0^{+4.9}_{-3.7}$ &	TS	& -0.28$^{+0.32}_{-0.30}$	& $1.85$ & $0.44$ & $1.56$ 	& 0.2 & -0.8 & $23.55\pm{0.09}$ 	& 1.0 \\
143	& J142601.7+352802\space\space\space14:26:01.71\space\space\space35:28:02.8	& 2.7\arcsec	& $17.8^{+7.8}_{-6.7}$ & $ 7.4^{+5.2}_{-4.0}$ & $10.4^{+6.5}_{-5.3}$ &	T	& 0.17$^{+0.37}_{-0.42}$	& $2.14$ & $0.33$ & $2.93$ 	& 0.8 & 0.3 & $23.11\pm{0.06}$ 	& 1.8 \\
144	& J142530.9+352756\space\space\space14:25:30.95\space\space\space35:27:56.8	& 2.4\arcsec	& $17.7^{+6.3}_{-5.1}$ & $14.6^{+5.3}_{-4.2}$ & $ 3.1^{+4.1}_{-2.9}$ &	TS	& -0.64$^{+0.33}_{-0.26}$	& $5.16$ & $1.58$ & $2.15$ 	& 0.6 & -0.9 & $20.75\pm{0.01}$ 	& 1.3 \\
145	& J142527.5+352656\space\space\space14:25:27.53\space\space\space35:26:56.2	& 3.1\arcsec	& $21.8^{+6.8}_{-5.6}$ & $18.3^{+5.8}_{-4.7}$ & $ 3.5^{+4.3}_{-3.1}$ &	TS	& -0.67$^{+0.29}_{-0.22}$	& $6.05$ & $1.90$ & $2.29$ 	& -0.3 & 0.9 & $17.44\pm{0.00}$ 	& 1.1 \\
146	& J142524.4+352542\space\space\space14:25:24.40\space\space\space35:25:42.1	& 2.5\arcsec	& $61.3^{+9.7}_{-8.6}$ & $39.2^{+7.8}_{-6.6}$ & $22.0^{+6.6}_{-5.4}$ &	TS	& -0.27$^{+0.15}_{-0.14}$	& $18.44$ & $4.42$ & $15.73$ 	& -0.7 & -0.4 & $21.50\pm{0.01}$ 	& 1.0 \\
147	& J142519.7+354432\space\space\space14:25:19.79\space\space\space35:44:32.2	& 2.1\arcsec	& $70.7^{+11.1}_{-10.0}$ & $50.7^{+9.0}_{-7.9}$ & $20.0^{+7.3}_{-6.2}$ &	TS	& -0.43$^{+0.15}_{-0.14}$	& $7.44$ & $2.01$ & $4.98$ 	& -0.5 & 0.4 & $21.89\pm{0.02}$ 	& 1.0 \\
148	& J142517.7+353754\space\space\space14:25:17.71\space\space\space35:37:54.3	& 1.1\arcsec	& $11.4^{+4.9}_{-3.7}$ & $ 1.2^{+2.8}_{-1.2}$ & $10.2^{+4.6}_{-3.4}$ &	T	& 0.79$^{+0.18}_{-0.37}$	& $1.40$ & $0.05$ & $2.92$ 	& 0.7 & -0.6 & $24.48\pm{0.17}$ 	& 1.3 \\
149	& J142516.0+354325\space\space\space14:25:16.02\space\space\space35:43:25.8	& 2.3\arcsec	& $17.6^{+8.2}_{-7.1}$ & $21.1^{+6.7}_{-5.5}$ & $ 0.0^{+5.6}_{-0.0}$ &	TS	& -1.00$^{+0.38}_{-0.00}$	& $1.43$ & $0.64$ & $<1.07$ 	& ... & ... & $>$25.7	& ...\\
150	& J142505.2+353729\space\space\space14:25:05.25\space\space\space35:37:29.5	& 1.6\arcsec	& $21.6^{+6.9}_{-5.7}$ & $19.3^{+6.1}_{-4.9}$ & $ 2.3^{+4.1}_{-2.3}$ &	TS	& -0.78$^{+0.29}_{-0.17}$	& $1.78$ & $0.60$ & $0.45$ 	& 0.2 & -0.2 & $>$25.7	& ...\\
151	& J142503.5+353859\space\space\space14:25:03.57\space\space\space35:38:59.2	& 1.7\arcsec	& $37.7^{+8.3}_{-7.2}$ & $22.6^{+6.3}_{-5.1}$ & $15.0^{+6.2}_{-5.0}$ &	TS	& -0.20$^{+0.21}_{-0.20}$	& $3.09$ & $0.70$ & $2.90$ 	& -0.6 & -0.2 & $20.26\pm{0.01}$ 	& 1.2 \\
152	& J142501.5+354000\space\space\space14:25:01.58\space\space\space35:40:00.3	& 1.9\arcsec	& $52.2^{+9.6}_{-8.5}$ & $35.0^{+7.6}_{-6.5}$ & $17.2^{+6.6}_{-5.4}$ &	TS	& -0.34$^{+0.18}_{-0.16}$	& $4.26$ & $1.07$ & $3.30$ 	& 1.0 & -1.6 & $23.70\pm{0.08}$ 	& 1.0 \\
153	& J142501.2+353057\space\space\space14:25:01.28\space\space\space35:30:57.4	& 2.2\arcsec	& $36.6^{+9.1}_{-8.0}$ & $15.1^{+6.1}_{-4.9}$ & $21.5^{+7.4}_{-6.3}$ &	TSH	& 0.18$^{+0.22}_{-0.24}$	& $3.54$ & $0.55$ & $4.90$ 	& -1.7 & 0.1 & $23.91\pm{0.08}$ 	& 1.0 \\
154	& J142459.2+353751\space\space\space14:24:59.24\space\space\space35:37:51.4	& 1.8\arcsec	& $35.4^{+8.5}_{-7.3}$ & $21.5^{+6.6}_{-5.4}$ & $13.8^{+6.1}_{-4.9}$ &	TS	& -0.21$^{+0.23}_{-0.22}$	& $2.89$ & $0.66$ & $2.66$ 	& -0.6 & -0.4 & $21.61\pm{0.02}$ 	& 1.2 \\
155	& J142611.8+353301\space\space\space14:26:11.85\space\space\space35:33:01.9	& 1.7\arcsec	& $20.5^{+7.2}_{-6.1}$ & $ 1.8^{+4.0}_{-1.8}$ & $18.7^{+6.6}_{-5.4}$ &	T	& 0.83$^{+0.15}_{-0.30}$	& $1.64$ & $0.05$ & $3.51$ 	& -0.4 & 0.6 & $22.63\pm{0.03}$ 	& 1.1 \\
156	& J142540.2+354623\space\space\space14:25:40.21\space\space\space35:46:23.0	& 3.8\arcsec	& $24.6^{+7.5}_{-6.4}$ & $ 5.3^{+4.3}_{-3.1}$ & $19.2^{+6.7}_{-5.5}$ &	T	& 0.57$^{+0.21}_{-0.27}$	& $6.70$ & $0.55$ & $12.41$ 	& -0.7 & 1.1 & $16.33\pm{0.00}$ 	& 1.6 \\
157	& J142539.8+354438\space\space\space14:25:39.87\space\space\space35:44:38.6	& 2.4\arcsec	& $26.3^{+8.3}_{-7.2}$ & $10.2^{+5.6}_{-4.4}$ & $16.1^{+6.8}_{-5.6}$ &	T	& 0.23$^{+0.27}_{-0.30}$	& $2.55$ & $0.37$ & $3.70$ 	& -1.3 & 0.4 & $19.91\pm{0.01}$ 	& 1.2 \\
158	& J142531.6+352659\space\space\space14:25:31.60\space\space\space35:26:59.0	& 3.5\arcsec	& $29.8^{+7.5}_{-6.4}$ & $ 4.7^{+4.1}_{-2.9}$ & $25.1^{+6.8}_{-5.6}$ &	TH	& 0.69$^{+0.16}_{-0.22}$	& $8.74$ & $0.52$ & $17.35$ 	& 2.1 & 2.0 & $24.71\pm{0.15}$ 	& 1.0 \\
159	& J142448.4+353227\space\space\space14:24:48.48\space\space\space35:32:27.4	& 2.8\arcsec	& $40.6^{+9.8}_{-8.7}$ & $26.4^{+7.3}_{-6.2}$ & $14.2^{+7.2}_{-6.1}$ &	T	& -0.29$^{+0.24}_{-0.23}$	& $5.80$ & $1.41$ & $4.83$ 	& -0.6 & 1.0 & $24.00\pm{0.14}$ 	& 1.1 \\
160	& J142447.8+353138\space\space\space14:24:47.80\space\space\space35:31:38.8	& 3.3\arcsec	& $22.2^{+9.0}_{-7.9}$ & $13.7^{+6.1}_{-4.9}$ & $ 8.5^{+7.2}_{-6.1}$ &	T	& -0.22$^{+0.38}_{-0.39}$	& $3.07$ & $0.71$ & $2.81$ 	& ... & ... & ...$^b$	& ...\\
161	& J142528.8+353342\space\space\space14:25:28.80\space\space\space35:33:42.0	& 0.6\arcsec	& $ 7.7^{+4.3}_{-3.1}$ & $ 5.8^{+3.8}_{-2.5}$ & $ 1.9^{+3.1}_{-1.7}$ &	S	& -0.50$^{+0.51}_{-0.36}$	& $0.57$ & $0.16$ & $0.33$ 	& 0.6 & -0.1 & $19.52\pm{0.00}$ 	& 1.3 \\
162	& J142558.2+353144\space\space\space14:25:58.29\space\space\space35:31:44.4	& 1.6\arcsec	& $12.3^{+5.9}_{-4.8}$ & $ 6.0^{+4.1}_{-2.9}$ & $ 6.3^{+4.9}_{-3.7}$ &	S	& 0.04$^{+0.41}_{-0.45}$	& $0.97$ & $0.18$ & $1.17$ 	& 0.5 & -0.2 & $25.08\pm{0.26}$ 	& 1.3 \\
163	& J142458.1+353919\space\space\space14:24:58.14\space\space\space35:39:19.4	& 2.5\arcsec	& $25.6^{+8.2}_{-7.1}$ & $16.9^{+6.2}_{-5.0}$ & $ 8.7^{+6.2}_{-5.0}$ &	S	& -0.32$^{+0.31}_{-0.29}$	& $2.58$ & $0.64$ & $2.06$ 	& 0.8 & 0.1 & $25.21\pm{0.27}$ 	& 0.9 \\
164	& J142535.9+353102\space\space\space14:25:35.96\space\space\space35:31:02.8	& 1.1\arcsec	& $14.4^{+5.7}_{-4.5}$ & $ 0.0^{+2.4}_{-0.0}$ & $15.3^{+5.6}_{-4.4}$ &	H	& 1.00$^{+0.00}_{-0.25}$	& $1.11$ & $<0.07$ & $2.78$ 	& -0.2 & -0.4 & $23.53\pm{0.08}$ 	& 1.5 \\
165	& J142527.4+353257\space\space\space14:25:27.46\space\space\space35:32:57.9	& 0.6\arcsec	& $ 8.6^{+4.5}_{-3.3}$ & $ 0.0^{+1.9}_{-0.0}$ & $ 9.8^{+4.5}_{-3.3}$ &	H	& 1.00$^{+0.00}_{-0.32}$	& $0.66$ & $<0.06$ & $1.76$ 	& 0.8 & -0.6 & $24.80\pm{0.31}$ 	& 1.3 \\
166	& J142545.4+352711\space\space\space14:25:45.44\space\space\space35:27:11.8	& 2.5\arcsec	& $19.8^{+7.2}_{-6.0}$ & $ 1.2^{+3.8}_{-1.2}$ & $18.6^{+6.6}_{-5.4}$ &	H	& 0.89$^{+0.10}_{-0.30}$	& $3.05$ & $0.07$ & $6.77$ 	& -1.9 & -0.0 & $23.48\pm{0.07}$ 	& 1.2 \\
167	& J142523.4+353512\space\space\space14:25:23.49\space\space\space35:35:12.4	& 0.7\arcsec	& $ 6.5^{+4.3}_{-3.1}$ & $ 0.0^{+2.4}_{-0.0}$ & $ 6.6^{+4.1}_{-2.9}$ &	H	& 1.00$^{+0.00}_{-0.55}$	& $0.50$ & $<0.07$ & $1.17$ 	& 0.1 & -0.4 & $20.80\pm{0.01}$ 	& 1.3 \\
168	& J142621.6+353931\space\space\space14:26:21.63\space\space\space35:39:31.1	& 3.2\arcsec	& $23.2^{+8.2}_{-7.0}$ & $ 1.6^{+4.1}_{-1.6}$ & $21.7^{+7.5}_{-6.4}$ &	H	& 0.87$^{+0.12}_{-0.28}$	& $2.71$ & $0.07$ & $5.96$ 	& 1.4 & -2.1 & $>$25.7	& ...\\
\enddata
\tablenotetext{a}{Source No. 22 is overlapped by bleeding of charge in the NDWFS $R$ image we used in this paper. Source position and $R$ band magnitude were measured from an older version of the NDWFS $R$ image.}
\tablenotetext{b}{Source No. 37 and 160 are overlapped by bleeding of charge in both the NDWFS $R$ and the older NDWFS $R$ images}
\end{deluxetable}

\end{document}